\documentclass[aps,prd,onecolumn,superscriptaddress,nofootinbib,floatfix]{revtex4-2}

\usepackage{amsmath,amssymb,bm}
\usepackage{graphicx}
\usepackage{slashed}
\usepackage{xcolor}
\usepackage[colorlinks=true,linkcolor=blue,citecolor=blue,urlcolor=purple]{hyperref}

\newcommand{\arxiv}[2][]{\href{https://arxiv.org/abs/#2}{arXiv:#2\if\relax\detokenize{#1}\relax\else\ \mbox{[#1]}\fi}}
\newcommand{\doilink}[2]{\href{https://doi.org/#1}{#2}}
\newcommand{\Lp}{\Lambda\to p\,\ell^-\bar\nu_\ell}
\newcommand{\Lpmu}{\Lambda\to p\,\mu^-\bar\nu_\mu}
\newcommand{\Lpe}{\Lambda\to p\,e^-\bar\nu_e}
\newcommand{\AFB}{A_{\rm FB}}
\newcommand{\Br}{\mathcal{B}}
\newcommand{\GeVsq}{\,\mathrm{GeV}^2}
\newcommand{\vecp}[1]{\vec{p}_{#1}}

\begin{document}

\title{Lepton polarization and sensitivity to the form factors in $\Lp$ with lattice QCD}

\author{Qazi Maaz Us Salam}
\email[Corresponding author: ]{qazimaaz92@gmail.com}
\affiliation{School of Science and Engineering, Lahore University of Management Sciences (LUMS), Opposite Sector U, D.H.A, Lahore 54792, Pakistan}
\affiliation{National Center for Physics, Shahdra Valley Road, Islamabad 44000, Pakistan}
\author{Ishtiaq Ahmed}
\email{ishtiaq.ahmed@ncp.edu.pk}
\affiliation{National Center for Physics, Shahdra Valley Road, Islamabad 44000, Pakistan}

\begin{abstract}
We investigate the polarization of the charged lepton in the semileptonic hyperon decay $\Lp$ ($\ell=e,\mu$) within the Standard Model, using the form factors obtained from lattice QCD. We compute the longitudinal, transverse and normal polarization of the lepton, the forward-backward asymmetry of each lepton spin state, and the lepton flavor universality ratio $R_{\mu e}$. We show that the longitudinal polarization determines the ratio $g_1/f_1$ of the axial-vector to vector couplings, and that the forward-backward asymmetries of the electron and of negative-helicity muons determine the weak magnetism. The asymmetry of positive-helicity muons is fixed by kinematics up to a small dependence on the second-class form factors, whereas $R_{\mu e}$ is nearly independent of $g_1/f_1$ and the weak magnetism. For comparison, we also use the QCD sum-rule form factors, whose $g_1/f_1$ and weak magnetism are opposite in sign to experiment and lattice QCD, and find very different values for the observables that depend on them. Moreover, the $q^2$ dependence of $R_{\mu e}$ fixes the longitudinal polarization of the muon, which can thus be measured from the decay rates, without measuring the muon spin. The normal polarization is generated by the electromagnetic final-state interaction and provides the Standard Model reference for searches for time-reversal violation. The electron forward-backward asymmetry and $R_{\mu e}$ can be measured at BESIII and the proposed Super Tau-Charm Facility, and $R_{\mu e}$ also at LHCb. These measurements can test the Standard Model and the lattice-QCD form factors.
\end{abstract}

\maketitle

\section{Introduction}
\label{sec:intro}

The semileptonic decays of hyperons offer valuable insight into the interplay of the weak and strong interactions in the light-baryon sector~\cite{Gaillard:1984ny,Garcia:1985xz,Cabibbo:2003cu}. Among these processes, the study of flavor changing charged current (FCCC) interactions, which occur at the quark level through the decay of a $s\to u\,\ell^-\bar\nu_\ell$. Such studies provide information about the Cabibbo-Kobayashi-Maskawa (CKM) matrix element $|V_{us}|$~\cite{Cabibbo:1963yz,Kobayashi:1973fv,Cabibbo:2003ea}, about the structure of the weak current, and about the breaking of the SU(3) flavor symmetry in baryonic form factors~\cite{Donoghue:1986th,Flores-Mendieta:1998tfv,Flores-Mendieta:2004cyh,Mateu:2005wi}. In this manuscript, we analyze in detail the decay $\Lp$ ($\ell=e,\mu$), which at the quark level corresponds to the transition $s\to u$, making it sensitive to the CKM matrix element $|V_{us}|$. Since the $\Lambda$ ($uds$) and the proton ($uud$) belong to the same baryon octet of the quark model~\cite{Gell-Mann:1964ewy}, this decay also serves as a testing ground for the SU(3) description of the baryon octet and its breaking.

The decay $\Lpe$ has a long experimental history. Its branching fraction and the ratio of the axial-vector to the vector coupling, $g_1/f_1$, were measured in the 1970s in bubble-chamber~\cite{Canter:1971zj} and spark-chamber~\cite{Althoff:1971hxl,Althoff:1971mul,Lindquist:1975fc} experiments, followed by the high-statistics experiments of Refs.~\cite{Wise:1980iq,Bristol-Geneva-Heidelberg-Orsay-Rutherford-Strasbourg:1983jzt,Dworkin:1990dd}, the last of which analyzed $37\,286$ $\Lambda$ $\beta$ decays. In a similar fashion, other hyperon semileptonic decays (HSDs) have been studied with the hyperon beam of the CERN SPS~\cite{Bristol-Geneva-Heidelberg-Orsay-Rutherford-Strasbourg:1983jzt,Bristol-Geneva-Heidelberg-Orsay-Rutherford-Strasbourg:1983jpz}, at Fermilab~\cite{Hsueh:1988ar,KTeVE832E799:1999tte,KTeV:2001djr}, and by the NA48/1 Collaboration~\cite{NA48I:2006yat,NA481:2012dtx}. The muon mode $\Lpmu$, whose branching fraction is about six times smaller, was studied already in the 1960s and 1970s~\cite{Ronne:1964zz,Canter:1971ay}; the 1971 measurement, for instance, was based on nine events~\cite{Canter:1971ay}.

The experimental situation for $\Lp$ has changed substantially in the last few years. Hyperon-antihyperon pairs produced in $J/\psi$ decays are polarized and their spins are entangled~\cite{Faldt:2017kgy,Perotti:2018wxm,BESIII:2018cnd}, which has opened a program of precision hyperon physics at BESIII~\cite{BESIII:2020nme,BESIII:2021ypr}. The formalism for semileptonic decays of such spin-entangled baryon-antibaryon pairs has been developed in Ref.~\cite{Batozskaya:2023rek}. BESIII measured the muon mode with tagged $\Lambda\bar\Lambda$ pairs~\cite{BESIII:2021ynj} and searched for the decay $\Xi^-\to\Xi^0e^-\bar\nu_e$~\cite{BESIII:2021emv}. Furthermore, LHCb whose prospects for hyperon physics are discussed in Ref.~\cite{AlvesJunior:2018ldo}, has since determined the branching fraction of the muon mode with twice the precision, $\Br(\Lpmu)=(1.462\pm0.016\pm0.100\pm0.011)\times10^{-4}$, and obtained $R_{\mu e}=0.175\pm0.012$~\cite{LHCb:2025wld}. In the electron mode, BESIII exploited the polarization and quantum entanglement of $\Lambda\bar\Lambda$ pairs from $J/\psi$ decays. This gave the absolute branching fraction, the axial-vector and weak-magnetism couplings, and the first determination of the weak-electricity coupling~\cite{BESIII:2026ehc}. The proposed Super Tau-Charm Facility (STCF) is expected to increase these samples by more than two orders of magnitude~\cite{Achasov:2023gey,Zhou:2025skh}.

Motivated by these experimental developments, numerous theoretical studies have investigated the HSDs using different approaches, including SU(3) analyses of the experimental data and the $1/N_c$ expansion~\cite{Donoghue:1986th,Cabibbo:2003cu,Flores-Mendieta:1998tfv,Flores-Mendieta:2004cyh,Mateu:2005wi}, quark models~\cite{Schlumpf:1994fb,Faessler:2008ix,Lih:2026dcj,Ni:2026arb}, the chiral quark-soliton model~\cite{Ledwig:2008ku}, chiral perturbation theory~\cite{Krause:1990xc,Anderson:1993as,Villadoro:2006nj,Lacour:2007wm,Geng:2009ik,Ledwig:2014rfa,Shanahan:2015dka}, QCD sum rules (QCDSR)~\cite{Zhang:2024ick,Ahmadi:2025oal}, and lattice QCD~\cite{Guadagnoli:2006gj,Lin:2007ap,Sasaki:2008ha,Sasaki:2012ne,Sasaki:2017jue,Bacchio:2025auj}, etc. The computation of the form factors requires the use of such non-perturbative approaches. The ratio $R_{\mu e}=\Gamma(\Lpmu)/\Gamma(\Lpe)$ has been studied in an effective-field-theory framework~\cite{Chang:2014iba} and computed in QCDSR~\cite{Ahmadi:2025oal}, in quark models~\cite{Ni:2026arb,Lih:2026dcj}, and in lattice QCD~\cite{Bacchio:2025auj}.

On the theory side, a lattice-QCD calculation of all six vector and axial-vector form factors of $\Lp$ with physical quark masses has recently been performed~\cite{Bacchio:2025auj}. It gives $g_1/f_1=0.6902(44)$, in agreement with the world average $0.718\pm0.015$~\cite{ParticleDataGroup:2026mpi} and the new BESIII value $0.729^{+0.048}_{-0.047}\pm0.007$~\cite{BESIII:2026ehc}. It also determines the weak-magnetism, second-class ($f_3$ and $g_2$) and induced-pseudoscalar form factors, which experiment constrains only weakly or not at all. Combined with the BESIII measurement, it yields $|V_{us}|=0.2339\pm0.0041$ from a hyperon decay~\cite{BESIII:2026ehc}. This is of interest in view of the present tensions in the first-row unitarity test of the CKM matrix, often referred to as the Cabibbo angle anomaly~\cite{Seng:2018yzq,Grossman:2019bzp,Cirigliano:2022yyo,FlavourLatticeAveragingGroupFLAG:2024oxs}. With these lattice form factors, observables beyond the branching ratio can now be predicted with good precision. The opportunities offered by semileptonic hyperon decays at present and future facilities are reviewed in Ref.~\cite{Adlarson:2026lsp}.

Semileptonic decays of light quarks are also sensitive probes of new physics (NP), in particular of non-standard scalar, pseudoscalar and tensor interactions~\cite{Cirigliano:2009wk,Bhattacharya:2011qm,Cirigliano:2013xha,Gonzalez-Alonso:2018omy,Falkowski:2020pma}. For $s\to u$ transitions, such analyses have been performed for kaon~\cite{Gonzalez-Alonso:2016etj} and hyperon~\cite{Chang:2014iba} decays. The chirality structure of new vector-like charged-current interactions has been discussed for heavy-meson decays in Ref.~\cite{Zafar:2025mhe}. In $s\to u$ transitions, muon-electron universality is tested most precisely in kaon decays, at the $0.5\%$ level in $K_{\ell3}$~\cite{FlaviaNetWorkingGrouponKaonDecays:2010lot}, and in the ratio of the leptonic decay rates $K\to e\nu$ and $K\to\mu\nu$~\cite{NA62:2012lny}. Hyperon decays provide an independent test of lepton flavor universality (LFU) with different hadronic input.

Investigating polarization effects in semileptonic decays provides a mechanism for exploring the weak interactions, provide information that cannot be obtained from decay rates alone. In particular, the polarization of the charged lepton has been studied extensively for the $\tau$ lepton in semileptonic $B$-meson decays~\cite{Tanaka:2010se,Ivanov:2017mrj,Alonso:2017ktd,Asadi:2020fdo} and in $\Lambda_b\to\Lambda_c\tau^-\bar\nu_\tau$~\cite{Gutsche:2015mxa}, and the $\tau$ polarization has been measured by the Belle Collaboration~\cite{Belle:2016dyj}. Lepton polarization asymmetries and polarized forward-backward asymmetries have also been proposed as probes of the rare decays $B\to K_1\ell^+\ell^-$~\cite{Ishaq:2013toa,Munir:2015gsp} and $\Lambda_b\to\Lambda^{(\ast)}\ell^+\ell^-$~\cite{Khan:2025rgc,Das:2026ikt}, and the observables of longitudinally and transversely polarized final-state mesons have been studied in $b\to c\,\ell\bar\nu$ transitions~\cite{Khan:2026mwk}. For baryons, the helicity formalism including lepton-mass effects~\cite{Korner:1989qb,Korner:1994nh} has been applied to heavy-baryon decays (see, e.g., Refs.~\cite{Gutsche:2014zna,Faustov:2016pal,Gutsche:2015rrt}) and, in our previous work, to the polarization observables of $\Xi_{cc}^{++}\to\Xi_c^+\bar\ell\nu_\ell$~\cite{Salam:2026qzj}. Lepton polarization in HSDs was analyzed with the helicity method in Ref.~\cite{Kadeer:2005aq}, with numerical results for $\Xi^0\to\Sigma^+\ell^-\bar\nu_\ell$.

In the muon mode, the phase space is very narrow, $q^2_{\max}\simeq2.8\,m_\mu^2$, so the helicity-flip amplitudes, which produce a lepton of positive helicity and vanish for $m_\ell\to0$, are of the same order as the non-flip ones. Recent studies of $\Lp$ in the light-front quark model~\cite{Lih:2026dcj} and in a relativistic three-quark model~\cite{Ni:2026arb} cover the form factors, the rates and, in the latter case, the lepton-antineutrino correlation and the asymmetries of a polarized $\Lambda$. However, the polarization of the charged lepton itself is not considered in any of these works.

Furthermore, the component of the lepton polarization vector normal to the decay plane is T-odd and, in the Standard Model (SM), vanishes at tree level. The muon polarization normal to the decay plane in $K_{\mu3}$ decays was proposed long ago as a test of time-reversal invariance~\cite{Sakurai:1958zza}, and it was soon realized that the electromagnetic final-state interaction (FSI) also generates such T-odd correlations, which form a background to the effects of CP violation~\cite{Khriplovich:1967sop,Okun:1967ww,Zhitnitsky:1980he,Efrosinin:2000yv}. This polarization has been searched for in $K^+\to\pi^0\mu^+\nu_\mu$ decays by the KEK-E246 Collaboration~\cite{KEK-E246:2004xuu,Abe:2006de}. T-odd correlations in HSDs are usually built from the spin of a baryon, for example $(\vecp{\ell}\times\vecp{B})\cdot\vec S_{B'}$~\cite{Adlarson:2026lsp}. The normal polarization of the charged lepton requires neither a polarized hyperon nor a measurement of the baryon spin. In the SM, it is generated by the electromagnetic FSI of the two charged particles, the proton and the muon. We compute this effect for $\Lp$. For hyperons, such FSI-induced T-odd correlations were considered early on for $\Xi^-\to\Lambda e^-\bar\nu_e$~\cite{Khriplovich:1967sop}.

In this study, we compute the longitudinal, transverse and normal polarization of the charged lepton in $\Lp$, together with the forward-backward asymmetry of each lepton spin state. Our main input is the set of lattice-QCD form factors of Ref.~\cite{Bacchio:2025auj}, whose statistical uncertainties and correlations we propagate together with an estimate of the systematic uncertainties (Sec.~\ref{sec:lattice}). For comparison, we use the form factors of the recent QCDSR analysis of Ref.~\cite{Ahmadi:2025oal}, which reproduce the measured branching ratios but whose ratio $g_1/f_1$ differs from experiment. This comparison shows which observables are fixed mainly by the lepton-side helicity structure and which are sensitive probes of the hadronic form factors. Furthermore, since BESIII now measures the axial, weak-magnetism and weak-electricity couplings directly~\cite{BESIII:2026ehc}, we determine which of these hadronic quantities each lepton observable measures. We find that the observables separate according to the hadronic quantity they measure. The longitudinal polarization measures the axial coupling, the electron forward-backward asymmetry and the asymmetry of negative-helicity muons measure the weak magnetism, and the asymmetry of positive-helicity muons measures only the second-class form factors. In addition, we compute the LFU ratio $R_{\mu e}$, which is nearly independent of the axial and weak-magnetism couplings. This makes it a clean SM prediction that complements the polarization observables. In the SM, its $q^2$ dependence, on the other hand, is fixed by the longitudinal polarization of the muon, which can therefore be determined from the rates, without measuring the muon spin. We use \textsc{FeynCalc}, a \textsc{Mathematica} package~\cite{Mertig:1990an,Shtabovenko:2016sxi,Shtabovenko:2020gxv,Shtabovenko:2023idz}, to compute the polarized squared amplitudes analytically, as in our previous works~\cite{Salam:2024nfv,Salam:2026qzj}.

We now give the structure of the rest of this paper. In Sec.~\ref{sec:theory}, we present the theoretical framework and define the polarization observables. In Sec.~\ref{sec:inputs}, we give the input parameters and the form factors with their uncertainties. In Sec.~\ref{sec:results}, we discuss the phenomenological analysis. In Sec.~\ref{sec:pn}, we compute the normal polarization. In Sec.~\ref{sec:ffsens}, we discuss the sensitivity of the observables to the form factors, and in Sec.~\ref{sec:concl}, we summarize our conclusions. The helicity expressions of the observables of Sec.~\ref{sec:results} are given in Appendix~\ref{app:helicity}, and the parametrizations of the QCDSR form factors in Appendix~\ref{app:qcdsr}.

\section{Theoretical framework}
\label{sec:theory}

In this section, we outline the theoretical framework that we use to compute the polarized observables of the decay $\Lp$. We first write the effective Hamiltonian and the parameterization of the hadronic matrix element in terms of form factors. We then define the kinematics and the spin projection of the lepton, and finally the physical observables of interest.

\begin{figure}[!h]
\centering
\includegraphics{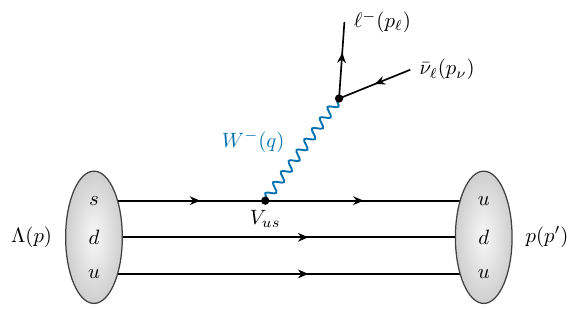}
\caption{Feynman diagram for the decay $\Lp$ in the SM.}
\label{fig:feynman}
\end{figure}

\subsection{Effective Hamiltonian and form factors}

At the quark level, the decay is governed by the $s\to u\,\ell^-\bar\nu_\ell$ transition. In the SM, it proceeds through the exchange of a virtual $W^-$ boson, as shown in Fig.~\ref{fig:feynman}, while the $u$ and $d$ quarks of the $\Lambda$ are spectators. Since the momentum transfer is negligible compared with the $W$ mass, $q^2\ll m_W^2$, the $W$ propagator shrinks to a point, and the transition is described by the effective Hamiltonian
\begin{equation}
\mathcal{H}_{\rm eff}=\frac{G_F}{\sqrt2}\,S_{\rm EW}^{1/2}\,V_{us}\,\big[\bar u\gamma^\mu(1-\gamma_5)s\big]\big[\bar\ell\gamma_\mu(1-\gamma_5)\nu_\ell\big]+\text{h.c.},
\label{eq:heff}
\end{equation}
where $G_F$ is the Fermi constant from muon decay, $V_{us}$ is the relevant CKM matrix element, and $S_{\rm EW}$ is the universal short-distance (SD) electroweak correction. We include $S_{\rm EW}$ in the effective Hamiltonian as in Eq.~(2.1) of Ref.~\cite{Cirigliano:2011ny}; its value is given in Eq.~(\ref{eq:sew}). The decay amplitude for $\Lambda(p)\to p(p')\,\ell^-(p_\ell)\,\bar\nu_\ell(p_\nu)$ can then be written as
\begin{equation}
\mathcal{M}=\frac{G_F}{\sqrt2}\,S_{\rm EW}^{1/2}\,V_{us}\,\langle p(p')|\bar u\gamma^\mu(1-\gamma_5)s|\Lambda(p)\rangle\;
\bar u_\ell(p_\ell)\gamma_\mu(1-\gamma_5)v_\nu(p_\nu),
\label{eq:amp}
\end{equation}
with $q=p-p'=p_\ell+p_\nu$. The hadronic current has the standard $V-A$ structure. For a $\frac12^+\to\frac12^+$ baryon transition, Lorentz covariance allows the hadronic matrix element to be decomposed in terms of six independent form factors. We use the standard decomposition~\cite{Weinberg:1958ut,Bacchio:2025auj},
\begin{equation}
\langle p(p')|\bar u\,\gamma^\mu(1-\gamma_5)\,s|\Lambda(p)\rangle=\bar u_p(p')\Big[\gamma^\mu\big(f_1-g_1\gamma_5\big)-\frac{i\sigma^{\mu\nu}q_\nu}{m_\Lambda}\big(f_2-g_2\gamma_5\big)+\frac{q^\mu}{m_\Lambda}\big(f_3-g_3\gamma_5\big)\Big]u_\Lambda(p),
\label{eq:ff}
\end{equation}
where $f_i\equiv f_i(q^2)$ and $g_i\equiv g_i(q^2)$. Here $f_1$ and $g_1$ are the vector and axial couplings, $f_2$ is the weak magnetism, $g_3$ is the induced pseudoscalar, and $f_3$, $g_2$ are the second-class form factors, which vanish in the SU(3) limit. The QCDSR analysis of Ref.~\cite{Ahmadi:2025oal} uses instead the basis $\{\gamma^\mu,\,p^\mu/m_\Lambda,\,p'^\mu/m_p\}$, with form factors $F_i$ and $G_i$. The two bases are related by the Gordon identities
\begin{align}
\bar u_p(p')(p+p')^\mu u_\Lambda(p)&=\bar u_p(p')\big[(m_\Lambda+m_p)\gamma^\mu+i\sigma^{\mu\nu}q_\nu\big]u_\Lambda(p),\nonumber\\
\bar u_p(p')(p+p')^\mu\gamma_5 u_\Lambda(p)&=\bar u_p(p')\big[(m_p-m_\Lambda)\gamma^\mu+i\sigma^{\mu\nu}q_\nu\big]\gamma_5 u_\Lambda(p),
\end{align}
which give, with $r\equiv m_p/m_\Lambda$,
\begin{equation}
F_1=f_1+(1+r)f_2,\quad F_2=f_3-f_2,\quad F_3=-r(f_2+f_3),\qquad
G_1=g_1-(1-r)g_2,\quad G_2=g_3-g_2,\quad G_3=-r(g_2+g_3).
\label{eq:basis}
\end{equation}
We have computed the squared amplitudes in the $\{F_i,G_i\}$ basis and use Eq.~(\ref{eq:basis}) to insert the lattice form factors.

\subsection{Kinematics}

\begin{figure}[!h]
\centering
\includegraphics{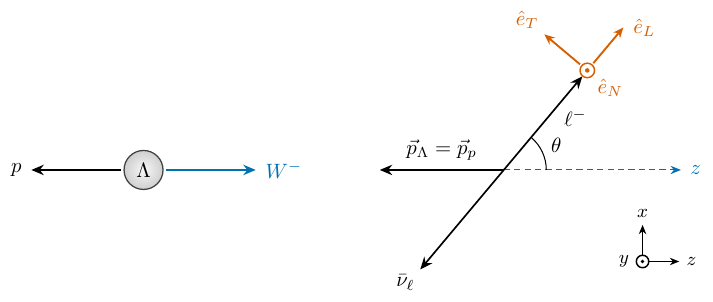}
\caption{Kinematics of $\Lp$ in the $\Lambda$ rest frame (left) and in the dilepton rest frame (right).}
\label{fig:kin}
\end{figure}

We work in the dilepton rest frame, where $\vecp{\Lambda}=\vecp{p}$, since $\vec q=0$. As shown in Fig.~\ref{fig:kin}, the proton and the virtual $W^-$ are emitted back to back in the $\Lambda$ rest frame, and the $\ell^-$ and $\bar\nu_\ell$ are emitted back to back in the dilepton rest frame. We define $\theta$ as the angle between the $\ell^-$ momentum and the direction opposite to $\vecp{p}$, i.e., the direction of the virtual $W^-$ in the $\Lambda$ rest frame. We choose the $z$ axis along this direction and the decay plane as the $xz$ plane, so that the $\Lambda$ and the proton move along $-z$ and $\theta$ is the polar angle of the $\ell^-$. The four-momenta then read
\begin{align}
p^\mu&=\big(E_\Lambda,\,0,\,0,\,-|\vecp{\Lambda}|\big), & p'^\mu&=\big(E_p,\,0,\,0,\,-|\vecp{\Lambda}|\big),\nonumber\\
p_\ell^\mu&=\big(E_\ell,\,|\vecp{\ell}|\sin\theta,\,0,\,|\vecp{\ell}|\cos\theta\big), & p_\nu^\mu&=|\vecp{\ell}|\,\big(1,\,-\sin\theta,\,0,\,-\cos\theta\big),
\label{eq:momenta}
\end{align}
with
\begin{equation}
E_{\Lambda,p}=\frac{m_\Lambda^2-m_p^2\pm q^2}{2\sqrt{q^2}},\qquad |\vecp{\Lambda}|=\frac{\lambda^{1/2}}{2\sqrt{q^2}},\qquad E_\ell=\frac{q^2+m_\ell^2}{2\sqrt{q^2}},\qquad |\vecp{\ell}|=\frac{q^2-m_\ell^2}{2\sqrt{q^2}},
\label{eq:energies}
\end{equation}
where $\lambda\equiv\lambda(m_\Lambda^2,m_p^2,q^2)$ is the Kallen function and $m_\ell^2\le q^2\le(m_\Lambda-m_p)^2$. With $x\equiv\cos\theta$, the scalar products that enter the squared amplitudes read
\begin{align}
p\cdot p_\ell&=\frac{1}{4q^2}\Big[(m_\Lambda^2-m_p^2+q^2)(q^2+m_\ell^2)+(q^2-m_\ell^2)\sqrt\lambda\,x\Big],\nonumber\\
p'\cdot p_\ell&=\frac{1}{4q^2}\Big[(m_\Lambda^2-m_p^2-q^2)(q^2+m_\ell^2)+(q^2-m_\ell^2)\sqrt\lambda\,x\Big].
\end{align}
It is important to mention here that the sign of the forward-backward asymmetries depends on this choice of $\theta$, which should be kept in mind when comparing with other conventions.

\subsection{Lepton spin projection}

To discuss the polarization of the charged lepton, we apply the spin projection operator $\frac12(1+\gamma_5\slashed{S})$ on the lepton spinor, by making the replacement
\begin{equation}
u_\ell(p_\ell)\bar u_\ell(p_\ell)\;\longrightarrow\;(\slashed{p}_\ell+m_\ell)\,\frac{1+\gamma_5\slashed{S}}{2}
\label{eq:projector}
\end{equation}
in the spin sum of the lepton. The spin four-vector satisfies $S^2=-1$ and $S\cdot p_\ell=0$. In the lepton rest frame $S=(0,\hat e_n)$, where we define a set of mutually orthogonal spin-quantization axes,
\begin{equation}
\hat e_L=\frac{\vecp{\ell}}{|\vecp{\ell}|},\qquad \hat e_N=\frac{\vecp{\ell}\times\vecp{\Lambda}}{|\vecp{\ell}\times\vecp{\Lambda}|},\qquad \hat e_T=\hat e_N\times\hat e_L ,
\end{equation}
and the labels $L$, $N$ and $T$ correspond to the longitudinal, normal and transverse polarization directions, respectively. Thus $\hat e_T$ lies in the decay plane, is perpendicular to the lepton momentum and has a positive projection on $\vecp{\Lambda}$. In the dilepton rest frame, the longitudinal and transverse spin vectors for spin projection $k=\pm1$ are
\begin{equation}
S_L^\mu=k\left(\frac{|\vecp{\ell}|}{m_\ell},\,\frac{E_\ell\,\vecp{\ell}}{m_\ell|\vecp{\ell}|}\right),\qquad S_T^\mu=k\,(0,\hat e_T),
\end{equation}
with $E_\ell$ and $|\vecp{\ell}|$ given in Eq.~(\ref{eq:energies}). In the coordinates of Eq.~(\ref{eq:momenta}), $\hat e_L=(\sin\theta,0,\cos\theta)$, $\hat e_T=(\cos\theta,0,-\sin\theta)$ and $\hat e_N=(0,1,0)$, as shown in Fig.~\ref{fig:kin}. The spin vectors enter the traces only through
\begin{align}
p\cdot S_L&=\frac{k}{4m_\ell q^2}\Big[(m_\Lambda^2-m_p^2+q^2)(q^2-m_\ell^2)+\sqrt\lambda\,(q^2+m_\ell^2)\,x\Big],\nonumber\\
p'\cdot S_L&=\frac{k}{4m_\ell q^2}\Big[(m_\Lambda^2-m_p^2-q^2)(q^2-m_\ell^2)+\sqrt\lambda\,(q^2+m_\ell^2)\,x\Big],\nonumber\\
p\cdot S_T&=p'\cdot S_T=-k\,\frac{\sqrt\lambda}{2\sqrt{q^2}}\,\sqrt{1-x^2}.
\end{align}
For the normal spin vector $S_N=k\,(0,\hat e_N)$, all scalar products with the momenta vanish, because the normal direction is perpendicular to the decay plane. The two normal spin states can then differ only through a term $\epsilon^{\mu\nu\rho\sigma}p_\mu p'_\nu p_{\ell\rho}S_{N\sigma}$ multiplied by an imaginary part, which is absent in the SM at tree level with real form factors. The normal polarization $P_N$ is generated at $\mathcal{O}(\alpha)$ by the FSI and is discussed in Sec.~\ref{sec:pn}.

The polarized squared amplitudes $\sum|\mathcal{M}_k|^2$, summed over the baryon and antineutrino spins with the projector~(\ref{eq:projector}) for the lepton, were computed analytically with \textsc{FeynCalc}~\cite{Mertig:1990an,Shtabovenko:2016sxi,Shtabovenko:2020gxv,Shtabovenko:2023idz}, keeping the full dependence on $x$. 

\subsection{Physical observables}

The polarized decay rate is given by
\begin{equation}
\frac{d^2\Gamma_k}{dq^2\,dx}=\frac{|\vecp{p}|_\Lambda}{512\,\pi^3 m_\Lambda^2}\left(1-\frac{m_\ell^2}{q^2}\right)\sum|\mathcal{M}_k|^2,\qquad
|\vecp{p}|_\Lambda=\frac{\lambda^{1/2}(m_\Lambda^2,m_p^2,q^2)}{2m_\Lambda},
\label{eq:rate}
\end{equation}
where $\sum|\mathcal{M}_k|^2$ is summed over the baryon and antineutrino spins, and the factor $\frac12$ for the average over the initial $\Lambda$ spin is included in the prefactor. Summing over $k$ gives the unpolarized rate, $d\Gamma=d\Gamma_++d\Gamma_-$, and the branching ratio can be written as $d\Br_k/dq^2=\tau_\Lambda\,d\Gamma_k/dq^2$. It is useful to study the polarization asymmetry along the direction $n\in\{L,T\}$, defined via
\begin{equation}
P_n(q^2)=\frac{d\Gamma_+^{(n)}/dq^2-d\Gamma_-^{(n)}/dq^2}{d\Gamma_+^{(n)}/dq^2+d\Gamma_-^{(n)}/dq^2}.
\end{equation}
The quantity $P_n(q^2)$ measures the polarization of the lepton along the direction $n$ through the normalized difference between the decay rates of its positive and negative spin states. The forward-backward asymmetry of leptons in the spin state $k$ is defined as
\begin{equation}
\AFB^{k}(q^2)=\frac{\int_0^1dx\,\frac{d^2\Gamma_k}{dq^2dx}-\int_{-1}^0dx\,\frac{d^2\Gamma_k}{dq^2dx}}{\int_{-1}^1dx\,\frac{d^2\Gamma_k}{dq^2dx}} .
\label{eq:afb}
\end{equation}
Because each spin state is normalized to its own rate, the unpolarized asymmetry is the rate-weighted average,
\begin{equation}
\AFB(q^2)=\frac{d\Gamma_+}{d\Gamma}\,\AFB^{+}(q^2)+\frac{d\Gamma_-}{d\Gamma}\,\AFB^{-}(q^2),
\label{eq:afbsum}
\end{equation}
for either spin direction. We also compute the $q^2$-averaged quantities, which are obtained by integrating the numerator and the denominator separately over $m_\ell^2\le q^2\le(m_\Lambda-m_p)^2$, e.g.\ $\langle P_n\rangle=(\Br_+^{(n)}-\Br_-^{(n)})/(\Br_+^{(n)}+\Br_-^{(n)})$.

It is worth mentioning here that experiments combine $\Lp$ with the charge-conjugate mode $\bar\Lambda\to\bar p\,\ell^+\nu_\ell$. For the latter, we define $\theta$, $\hat e_L$, $\hat e_N$ and $\hat e_T$ in the same way, with $\Lambda$, $p$, $\ell^-$ replaced by $\bar\Lambda$, $\bar p$, $\ell^+$. In the SM the decay conserves CP. Under CP all momenta change sign while the spins do not, so $\theta$ and $\hat e_N$ are unchanged, whereas $\hat e_L$ and $\hat e_T$ change sign. Hence the rates, $R_{\mu e}$ and the unpolarized $\AFB$ are the same in both modes. The two spin states along $\hat e_L$ and along $\hat e_T$ are interchanged, so $P_L$ and $P_T$ change sign and $\AFB^{\pm}$ of one mode equals $\AFB^{\mp}$ of the other. $P_N$ has the same sign in both modes, unless CP is violated. In particular, a massless $\ell^+$ has $P_L=+1$. We have verified these relations by computing $\bar\Lambda\to\bar p\,\ell^+\nu_\ell$ directly, with the antibaryon current obtained from the charge-conjugation invariance of QCD and the $\ell^+$ spin projector $(\slashed p_\ell-m_\ell)(1+\gamma_5\slashed S)/2$.

\section{Input parameters and form factors}
\label{sec:inputs}

In this section, we list the numerical input parameters, and we discuss the form factors used in our analysis together with their uncertainties.

\subsection{Input parameters}

We show, in Table~\ref{tab:inputs}, various parameters (masses, the Fermi constant, the relevant CKM matrix element and the lifetime of the $\Lambda$) used in our analysis~\cite{ParticleDataGroup:2026mpi}.

\begin{table}[h]
\caption{Numerical values of various input parameters used in our numerical analysis~\cite{ParticleDataGroup:2026mpi}.}
\label{tab:inputs}
\centering
\begin{tabular}{l@{\hspace{1.5em}}l@{\hspace{4em}}l@{\hspace{1.5em}}l}
\hline\hline\noalign{\vskip3pt}
$m_\Lambda$ & $1115.68$ MeV & $\tau_\Lambda$ & $2.617\times10^{-10}$ s\\
$m_p$ & $938.27$ MeV & $G_F$ & $1.16638\times10^{-5}\,\mathrm{GeV}^{-2}$\\
$m_\mu$ & $105.66$ MeV & $|V_{us}|$ & $0.2243$\\
$m_e$ & $0.511$ MeV & & \\[3pt]
\hline\hline
\end{tabular}
\end{table}

The Fermi constant measured in muon decay does not contain the SD electroweak corrections specific to semileptonic decays of hadrons. These enhance all semileptonic rates by a universal factor~\cite{Marciano:2005ec,Cirigliano:2011ny}. We include its leading logarithms in the form given by Marciano and Sirlin for baryon $\beta$ decay~\cite{Marciano:2005ec},
\begin{equation}
S_{\rm EW}=1+\frac{\alpha}{2\pi}\left[3\ln\frac{M_Z}{m_p}+\ln\frac{M_Z}{m_A}\right]=1.0210 ,
\label{eq:sew}
\end{equation}
where $\alpha=1/137.036$, $M_Z=91.1876$~GeV, and $m_A=1.2$~GeV is the low-energy cutoff of the axial-vector $\gamma W$ box. The first logarithm comes from the vector-current loops and the second from the axial-vector $\gamma W$ box. Two equivalent forms give practically the same value: the leading-logarithmic expression $S_{\rm EW}=1+(2\alpha/\pi)\ln(M_Z/\mu)$~\cite{Cirigliano:2011ny} gives $1.0213$ for $\mu=m_p$, and the factor $(1+0.0105)^2$ used in Ref.~\cite{Bacchio:2025auj} gives $1.0211$.

\subsection{Lattice-QCD form factors }
\label{sec:lattice}

The hadronization of quarks and gluons is described in terms of form factors that are the main source of theoretical uncertainties. Our main input is the lattice-QCD calculation of Ref.~\cite{Bacchio:2025auj}, performed on a single ensemble with physical light, $s$ and $c$ quark masses. Since $q^2\le0.0315\GeVsq$, the form factors are well described by their values and slopes at $q^2=0$,
\begin{equation}
f_i(q^2)=f_i\Big(1+\frac{\langle r^2_{f_i}\rangle}{6}\,q^2\Big),\qquad g_i(q^2)=g_i\Big(1+\frac{\langle r^2_{g_i}\rangle}{6}\,q^2\Big),
\end{equation}
which is accurate to the sub-percent level~\cite{Bacchio:2025auj}. Table~\ref{tab:lattice} lists the charges and radii in the basis of Eq.~(\ref{eq:ff}), with the radii in units of $1/m_\Lambda^2$; we also use the correlation matrices given in Ref.~\cite{Bacchio:2025auj}.

\begin{table}[h]
\caption{Lattice-QCD charges and radii of the $\Lambda\to p$ form factors~\cite{Bacchio:2025auj}.}
\label{tab:lattice}
\begin{ruledtabular}
\begin{tabular}{lcccccc}
 & $f_1$ & $f_2$ & $f_3$ & $g_1$ & $g_2$ & $g_3$\\ \hline
charge & $1.1849(57)$ & $0.821(19)$ & $0.062(13)$ & $0.8178(55)$ & $-0.082(19)$ & $-5.14(11)$\\
$\langle r^2\rangle\,m_\Lambda^2$ & $9.50(28)$ & $17.0(23)$ & $24.7(68)$ & $6.46(47)$ & $38(14)$ & $24.7(23)$\\
\end{tabular}
\end{ruledtabular}
\end{table}

We propagate the errors of Table~\ref{tab:lattice}, including correlations. Since they come from a single ensemble without continuum extrapolation, we add three systematic uncertainties in quadrature: an uncorrelated $2\%$ on the six charges for discretization and finite-volume effects, the shift from setting $g_1/f_1$ to its experimental value $0.718$~\cite{ParticleDataGroup:2026mpi}, and the shift from scaling all radii by $1.5$, a variation much larger than their lattice errors, for the $q^2$ shape. Table~\ref{tab:budget} lists these contributions for the muon channel. For the asymmetries, the statistical errors are small, and the total is dominated by the $2\%$ and $g_1/f_1$ terms. The dependence on the individual couplings is studied in Sec.~\ref{sec:ffsens}. All lattice predictions below are given with this total uncertainty.

\begin{table}[h]
\caption{Sources of uncertainty for the muon channel with the lattice form factors.}
\label{tab:budget}
\begin{ruledtabular}
\begin{tabular}{lcccccc}
 & $\Br_\mu\times10^4$ & $\langle P_L\rangle$ & $\langle P_T\rangle$ & $\langle\AFB\rangle$ & $\langle\AFB^{-}\rangle_L$ & $\langle\AFB^{-}\rangle_T$\\ \hline
Central value & $1.217$ & $-0.0807$ & $0.5286$ & $-0.1679$ & $0.0963$ & $-0.1314$\\
Statistical/fit & $0.010$ & $0.0036$ & $0.0021$ & $0.0013$ & $0.0015$ & $0.0021$\\
Statistical $\oplus$ 2\% on charges & $0.036$ & $0.0139$ & $0.0070$ & $0.0037$ & $0.0022$ & $0.0061$\\
$g_1/f_1\to0.718$ & $+0.058$ & $-0.0186$ & $-0.0099$ & $+0.0040$ & $-0.0023$ & $+0.0067$\\
Radii $\times1.5$ & $+0.029$ & $+0.0014$ & $+0.0007$ & $+0.0003$ & $+0.0007$ & $+0.0001$\\ \hline
Total & $0.074$ & $0.023$ & $0.012$ & $0.005$ & $0.003$ & $0.009$\\
\end{tabular}
\end{ruledtabular}
\end{table}

\subsection{QCD sum-rule form factors }
\label{sec:qcdsr}

For comparison, we use the QCDSR form factors of Ref.~\cite{Ahmadi:2025oal}, given in the basis $\{F_i,G_i\}$ with a polynomial and a $z$-expansion parametrization (Appendix~\ref{app:qcdsr}). Their values at $q^2=0$ are $F_{1,2,3}(0)=-0.24(8),\,-0.11(3),\,1.58(47)$ and $G_{1,2,3}(0)=-0.76^{+0.21}_{-0.22},\,-0.44(13),\,-1.50(45)$; the $q^2$-shape parameters are given in Ref.~\cite{Ahmadi:2025oal}. Converting with Eq.~(\ref{eq:basis}) gives the couplings in Table~\ref{tab:couplings}, where we also list the SU(3) values and the experimental results. In the SU(3) limit, $f_2/f_1=(m_\Lambda/m_p)\,\mu_p/2$, with $\mu_p$ the anomalous magnetic moment of the proton~\cite{Cabibbo:2003cu}. The lattice $f_1$ is only $3\%$ below its SU(3) value, and the lattice $g_1/f_1$ and $f_2/f_1$ agree with experiment. The lattice $f_2/f_1$ is, however, $35\%$ below its SU(3) value. This is expected: the Ademollo-Gatto theorem protects $f_1$, but not $f_2$, against first-order SU(3) breaking~\cite{Cabibbo:2003cu}. The QCDSR ratio $g_1/f_1$, on the other hand, differs from experiment in magnitude and in sign. We note that the sign of $g_1/f_1$ is physical once the vector and axial matrix elements are written in the same form, as in Eq.~(\ref{eq:ff}). The PDG quotes $g_A/g_V=-0.718\pm0.015$ for the current $\bar B_f\gamma_\lambda(g_V+g_A\gamma_5)B_i$~\cite{ParticleDataGroup:2026mpi}, which corresponds to $g_1/f_1=-g_A/g_V=+0.718\pm0.015$ in Eq.~(\ref{eq:ff}). The QCDSR vector coupling is dominated by $F_3(0)$: in $f_1=F_1+(m_\Lambda+m_p)\big[F_2/(2m_\Lambda)+F_3/(2m_p)\big]$ the $F_3$ term alone contributes $1.73$, so the large uncertainty of $F_3(0)$ dominates the uncertainties of the QCDSR predictions.

The QCDSR uncertainties are obtained by shifting each $F_i(0)$ and $G_i(0)$ by its upper and lower error, treated as uncorrelated, and adding the upward and downward deviations separately in quadrature. Uncertainties of the $q^2$-shape parameters are not available. We use the polynomial parametrization throughout, and show the $z$ expansion only in Tables~\ref{tab:br} and~\ref{tab:rmuebins} for comparison.

\begin{table}[h]
\caption{Vector, axial and weak-magnetism couplings at $q^2=0$.}
\label{tab:couplings}
\begin{ruledtabular}
\begin{tabular}{lcccc}
 & $f_1(0)$ & $g_1(0)$ & $g_1/f_1$ & $f_2/f_1$\\ \hline
Lattice QCD~\cite{Bacchio:2025auj} & $1.1849(57)$ & $0.8178(55)$ & $0.6902(44)$ & $0.693(17)$\\
QCD sum rules~\cite{Ahmadi:2025oal} & $1.39$ & $-0.58$ & $-0.42^{+0.19}_{-0.30}$ & $-0.64$\\
SU(3) limit~\cite{Cabibbo:2003cu} & $\sqrt{3/2}=1.225$ & & & $1.066$\\
Experiment, world average~\cite{ParticleDataGroup:2026mpi} & & & $0.718\pm0.015$ & \\
Experiment, BESIII, fit with $g_2=0$~\cite{BESIII:2026ehc} & & & $0.729^{+0.048}_{-0.047}\pm0.007$ & $0.89\pm0.35\pm0.14$\\
\end{tabular}
\end{ruledtabular}
\end{table}

\section{Phenomenological analysis}
\label{sec:results}

We now go on to present the phenomenological analysis of the observables defined in Sec.~\ref{sec:theory}. We begin with the unpolarized branching ratios and $R_{\mu e}$, and then discuss the longitudinal and transverse polarization of the muon, the $q^2$ dependence of $R_{\mu e}$ and the electron channel. We have used a consistent color scheme for all the figures: blue is used for the positive spin state, orange for the negative spin state, and black for the unpolarized result. Solid lines with bands correspond to the lattice-QCD form factors with their total uncertainty, while dashed lines show the QCDSR form factors (central values).

\subsection{Branching ratios}

We compare, in Table~\ref{tab:br}, the unpolarized branching ratios and the LFU ratio $R_{\mu e}=\Gamma(\Lpmu)/\Gamma(\Lpe)$, computed with the inputs of Table~\ref{tab:inputs}, with other predictions and measurements. With the lattice form factors, we obtain
\begin{equation}
R_{\mu e}=0.1665\pm0.0018 .
\label{eq:rmue}
\end{equation}
This agrees with the lattice result obtained with physical baryon masses, $0.16638(20)$~\cite{Bacchio:2025auj}. Our larger uncertainty includes the systematic variations of Sec.~\ref{sec:lattice}. The main lattice result, $0.1735(98)$~\cite{Bacchio:2025auj}, uses lattice baryon masses, which dominate its uncertainty. At next-to-leading order in SU(3) breaking, $R_{\mu e}$ is independent of the form factors~\cite{Chang:2014iba}. Evaluating Eq.~(9) of Ref.~\cite{Chang:2014iba} with the physical masses, we obtain $0.162$, $2.6\%$ below our value, as expected from the neglected terms of order $\delta^2$, $\delta=(m_\Lambda-m_p)/m_\Lambda$. The value $0.153(8)$ quoted in that work, and used for comparison in Refs.~\cite{BESIII:2021ynj,LHCb:2025wld}, is not reproduced by this expression. The measurements by BESIII, $0.178\pm0.028$~\cite{BESIII:2021ynj}, and LHCb, $0.175\pm0.012$~\cite{LHCb:2025wld}, agree with the SM. For the LHCb branching ratio in Table~\ref{tab:br}, the statistical, systematic and normalization uncertainties are combined in quadrature. We will show in Sec.~\ref{sec:ffsens} that $R_{\mu e}$ is nearly independent of the axial and weak-magnetism couplings.

The SD correction $S_{\rm EW}$ of Eq.~(\ref{eq:sew}) is the same for the electron and the muon, so it cancels in $R_{\mu e}$ and in all asymmetries. This is not the case for the long-distance QED corrections, which depend on the lepton mass. We do not include them in our predictions, but we estimate their size here. For HSDs, the model-independent radiative corrections to the lepton energy spectrum and the Dalitz plot have been obtained in Refs.~\cite{Garcia:1985xz,JuarezWysozka:1987si,Flores-Mendieta:1996vgq,Martinez:2000wp,Torres:2004zh,Flores-Mendieta:2004sab}, based on the general analysis of Ref.~\cite{Sirlin:1967zza}. For the muon polarization in $K_{\mu3}$ decays, the same model-independent approach gives corrections at the percent level or below~\cite{Sanchez-Glez:2020sgh,Vieyra:2023yil}. For $\Lpe$, they change the rate by $(1.6\pm0.5)\%$~\cite{Adlarson:2026lsp}; we are not aware of a calculation for the muon mode. For the muon mode, the Coulomb attraction between the proton and the muon is important, because both particles are slow. We estimate it with the Coulomb (Fermi) factor~\cite{Fermi:1934hr} $F=2\pi\eta/(1-e^{-2\pi\eta})$, $\eta=\alpha/\beta_{\rm rel}$, where $\beta_{\rm rel}=[1-m_p^2m_\ell^2/(p'\cdot p_\ell)^2]^{1/2}$ is the relative velocity of the proton and the lepton, averaged over the phase space. This factor raises $R_{\mu e}$ by $2.1\%$. The emission of real photons, which is larger for the electron, partly cancels this increase, as in $K^0_{\ell3}$ decays~\cite{Cirigliano:2011ny,FlaviaNetWorkingGrouponKaonDecays:2010lot}. We therefore expect the long-distance correction to raise $R_{\mu e}$ by roughly $1\%$ to $2\%$. This is larger than the uncertainty of Eq.~(\ref{eq:rmue}), so a complete calculation will be needed for the precision expected at future facilities.

For the branching ratios themselves, our differential rate agrees exactly with Eq.~(32) of Ref.~\cite{Bacchio:2025auj} at every $q^2$. With the SD factor of Eq.~(\ref{eq:sew}), our integrated rates are $\Gamma/|V_{us}|^2=5.551\times10^7\,{\rm s}^{-1}$ (electron) and $0.9240\times10^7\,{\rm s}^{-1}$ (muon). They agree to about $0.1\%$ with the values obtained in Ref.~\cite{Bacchio:2025auj} with physical baryon masses, $5.546(51)\times10^7\,{\rm s}^{-1}$ and $0.9228(82)\times10^7\,{\rm s}^{-1}$. The main results of Ref.~\cite{Bacchio:2025auj} listed in Table~\ref{tab:br} are higher because they use the lattice baryon masses. With $|V_{us}|=0.2243$, the electron-mode branching ratio, $(7.31\pm0.42)\times10^{-4}$, lies below the world average $(8.34\pm0.14)\times10^{-4}$~\cite{ParticleDataGroup:2026mpi} and the new BESIII value $(8.16\pm0.22\pm0.15)\times10^{-4}$~\cite{BESIII:2026ehc}. This is the known tension that, read in reverse, gives a larger $|V_{us}|$ from this decay: $0.2338(75)$ in Ref.~\cite{Bacchio:2025auj}, and $0.2339(41)$ when the BESIII measurement of the branching fraction and couplings is combined with the lattice form factors~\cite{BESIII:2026ehc}. It affects only the branching ratios, not the asymmetries discussed below.

With the QCDSR form factors, $G_F=1.17\times10^{-5}\,\mathrm{GeV}^{-2}$ and no $S_{\rm EW}$, as in Ref.~\cite{Ahmadi:2025oal}, and $|V_{us}|=0.218$, we reproduce the branching ratios of that work to better than $1\%$ for the polynomial fit.

\begin{table}[h]
\caption{Unpolarized branching ratios and $R_{\mu e}$.}
\label{tab:br}
\begin{ruledtabular}
\begin{tabular}{lccc}
 & $\Br(\Lpe)\times10^4$ & $\Br(\Lpmu)\times10^4$ & $R_{\mu e}$\\ \hline
This work, lattice form factors & $7.31\pm0.42$ & $1.217\pm0.074$ & $0.1665\pm0.0018$\\
This work, QCDSR form factors (polynomial) & $8.19^{+5.00}_{-3.48}$ & $1.62^{+1.02}_{-0.70}$ & $0.197^{+0.006}_{-0.008}$\\
This work, QCDSR form factors ($z$ expansion) & $7.77^{+4.40}_{-3.12}$ & $1.39^{+0.73}_{-0.53}$ & $0.179^{+0.006}_{-0.005}$\\
Lattice QCD, Ref.~\cite{Bacchio:2025auj} & $7.68(48)$ & $1.33(16)$ & $0.1735(98)$\\
QCD sum rules, Ref.~\cite{Ahmadi:2025oal} (polynomial) & $7.64^{+3.56}_{-1.25}$ & $1.50^{+0.80}_{-0.32}$ & $0.196^{+0.009}_{-0.012}$\\
QCD sum rules, Ref.~\cite{Zhang:2024ick} & $7.72\pm0.64$ & $1.35\pm0.11$ & \\
BESIII, Ref.~\cite{BESIII:2021ynj} & & $1.48\pm0.21\pm0.08$ & $0.178\pm0.028$\\
BESIII, Ref.~\cite{BESIII:2026ehc} & $8.16\pm0.22\pm0.15$ & & \\
LHCb, Ref.~\cite{LHCb:2025wld} & & $1.46\pm0.10$ & $0.175\pm0.012$\\
PDG, Ref.~\cite{ParticleDataGroup:2026mpi} & $8.34\pm0.14$ & $1.51\pm0.19$ & \\
\end{tabular}
\end{ruledtabular}
\end{table}

\subsection{Polarized observables of the muon}
\label{sec:muonpol}

\begin{figure}[t]
\centering
\includegraphics[width=\textwidth]{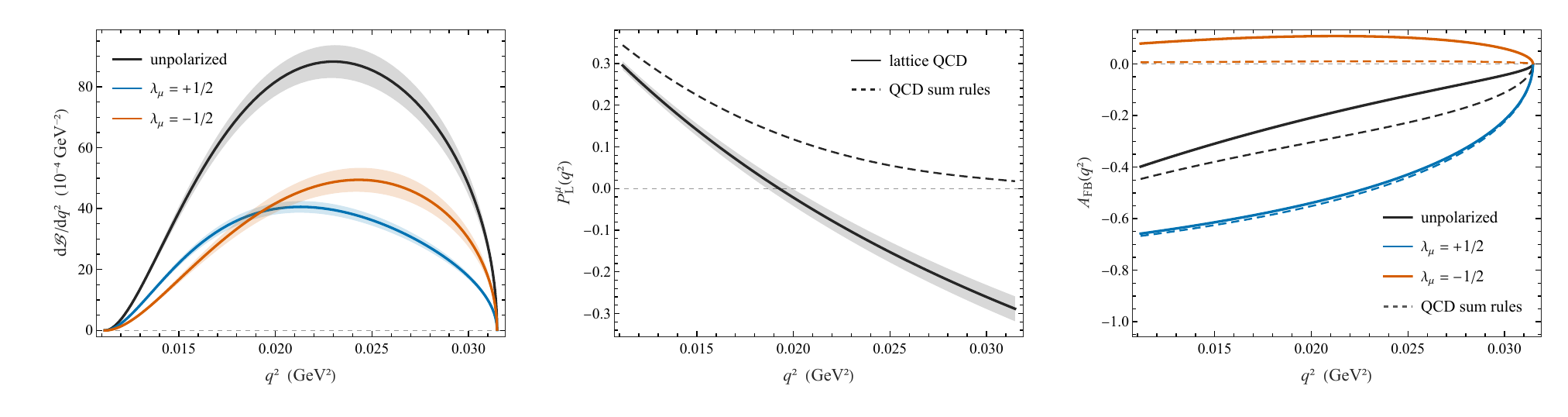}
\caption{Longitudinal polarization of the muon in $\Lpmu$: differential branching ratio (left), $P_L^\mu(q^2)$ (middle) and $\AFB(q^2)$ (right).}
\label{fig:long}
\end{figure}

In Fig.~\ref{fig:long}, we present the longitudinally polarized branching ratios, the polarization asymmetry $P_L^\mu(q^2)$ and the forward-backward $A_{\rm FB}(q^2)$ asymmetries of the muon, while the $q^2$-averaged values are collected in Table~\ref{tab:pol}. It can be seen from the left plot that the two muon helicities, $\lambda_\mu=\pm1/2$ (spin projection $k=\pm1$ along $\hat e_L$), contribute with comparable rates, $\Br_+^{(L)}=(0.559\pm0.027)\times10^{-4}$ and $\Br_-^{(L)}=(0.657\pm0.051)\times10^{-4}$. The helicity-flip ($\lambda_\mu=+1/2$) state dominates near threshold and the non-flip state dominates at large $q^2$. As a result, $P_L^\mu$ falls from $+0.30\pm0.01$ at threshold to $-0.29\pm0.03$ at $q^2_{\max}$, and it changes sign at $q_0^2=0.0193\pm0.0006\GeVsq$, where the flip and non-flip rates are equal, $\delta_\ell(H_U+H_L+3H_S)=H_U+H_L$ in the notation of Appendix~\ref{app:helicity}. The averaged polarization is small and negative, $\langle P_L^\mu\rangle=-0.081\pm0.023$. With the QCDSR form factors, on the other hand, $P_L^\mu(q^2)$ stays positive over the whole range and $\langle P_L^\mu\rangle=+0.095^{+0.200}_{-0.275}$. The zero crossing and the sign of the average therefore test the form factors, specifically the axial coupling $g_1/f_1$ (Sec.~\ref{sec:ffsens}).

The right plot of Fig.~\ref{fig:long} shows that the $A_{\rm FB}(q^2)$ depends strongly on the muon helicity. Positive-helicity muons are emitted preferentially in the backward direction, with $\AFB^{+}$ rising from $-0.658$ at threshold to zero at $q^2_{\max}$ and $\langle\AFB^{+}\rangle_L=-0.4785\pm0.0008$. This asymmetry is set almost entirely by the kinematics of the helicity-flip amplitude, and only the second-class form factors can move it (Sec.~\ref{sec:ffsens}). The QCDSR form factors give practically the same value, $-0.477$, but partly by accident: their large second-class couplings alone would shift the lattice value to $-0.525$, and their other differences shift it back. Negative-helicity muons, in contrast, are nearly isotropic. Their small asymmetry, $\langle\AFB^{-}\rangle_L=0.096\pm0.003$, comes from the parity-odd interference of the vector and axial transverse amplitudes, $\propto F_\perp G_\perp$, and depends mainly on the weak magnetism through $F_\perp$. The QCDSR form factors give $0.009$. The unpolarized asymmetry follows from Eq.~(\ref{eq:afbsum}), $\langle\AFB\rangle=-0.168\pm0.005$ (QCDSR: $-0.257^{+0.075}_{-0.054}$).

\begin{figure}[t]
\centering
\includegraphics[width=\textwidth]{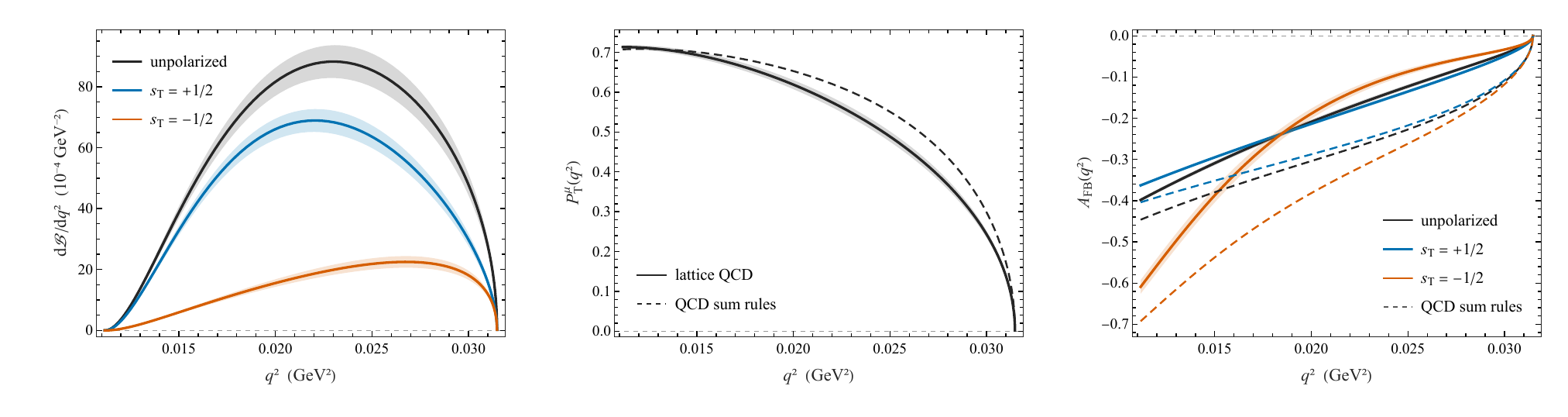}
\caption{Transverse polarization of the muon in $\Lpmu$: differential branching ratio (left), $P_T^\mu(q^2)$ (middle) and $\AFB(q^2)$ (right).}
\label{fig:trans}
\end{figure}

We now move on to the transverse polarization of the muon, which is shown in Fig.~\ref{fig:trans} for the two spin states $s_T=\pm1/2$ along $\hat e_T$. The muon is produced predominantly with its spin along $+\hat e_T$, i.e., tilted toward the direction of the baryons in the dilepton rest frame: $\Br_+^{(T)}=(0.930\pm0.051)\times10^{-4}$ against $\Br_-^{(T)}=(0.287\pm0.023)\times10^{-4}$. The asymmetry $P_T^\mu$ starts at $0.713\pm0.005$ at threshold and vanishes at $q^2_{\max}$. Its average, $\langle P_T^\mu\rangle=0.529\pm0.012$, is the largest polarization effect in the muon mode. It is also insensitive to the form factors: the QCDSR form factors give $0.569^{+0.112}_{-0.155}$, which is compatible within uncertainties. The forward-backward asymmetries of the two transverse spin states are $\langle\AFB^{+}\rangle_T=-0.179\pm0.004$ and $\langle\AFB^{-}\rangle_T=-0.131\pm0.009$ (QCDSR: $-0.249$ and $-0.287$). Moreover, when combined according to Eq.~(\ref{eq:afbsum}), they reproduce the unpolarized $\langle\AFB\rangle=-0.168$.

\begin{table}[h]
\caption{$q^2$-averaged observables of $\Lpmu$ for longitudinal ($L$) and transverse ($T$) spin quantization.}
\label{tab:pol}
\begin{ruledtabular}
\begin{tabular}{llccccc}
 & Form factors & $\Br_+\times10^4$ & $\Br_-\times10^4$ & $\langle P_n\rangle$ & $\langle\AFB^{+}\rangle$ & $\langle\AFB^{-}\rangle$\\ \hline
$L$ & lattice & $0.559\pm0.027$ & $0.657\pm0.051$ & $-0.081\pm0.023$ & $-0.4785\pm0.0008$ & $0.096\pm0.003$\\
$L$ & QCDSR & $0.885^{+0.694}_{-0.477}$ & $0.732^{+0.397}_{-0.280}$ & $0.095^{+0.200}_{-0.275}$ & $-0.477^{+0.017}_{-0.008}$ & $0.009^{+0.005}_{-0.004}$\\
$T$ & lattice & $0.930\pm0.051$ & $0.287\pm0.023$ & $0.529\pm0.012$ & $-0.179\pm0.004$ & $-0.131\pm0.009$\\
$T$ & QCDSR & $1.269^{+0.865}_{-0.597}$ & $0.349^{+0.189}_{-0.134}$ & $0.569^{+0.112}_{-0.155}$ & $-0.249^{+0.054}_{-0.034}$ & $-0.287^{+0.143}_{-0.160}$\\
\end{tabular}
\end{ruledtabular}
\end{table}

\subsection{The ratio $R_{\mu e}(q^2)$ and the muon helicity}
\label{sec:rmueq}

We now turn our attention to the ratio of the differential rates,
\begin{equation}
R_{\mu e}(q^2)=\frac{d\Gamma(\Lpmu)/dq^2}{d\Gamma(\Lpe)/dq^2},
\label{eq:rmuediff}
\end{equation}
which, as we show now, is fixed by the longitudinal polarization of the muon. Negative-helicity leptons ($\lambda_\ell=-1/2$) are produced by the non-flip amplitudes, which are the same for both leptons~\cite{Korner:1989qb,Kadeer:2005aq}, so that, neglecting the electron mass,
\begin{equation}
\frac{d\Gamma^{(L)}_-(\Lpmu)}{dq^2}=\Big(1-\frac{m_\mu^2}{q^2}\Big)^2\,\frac{d\Gamma(\Lpe)}{dq^2}.
\label{eq:nonflip}
\end{equation}
With $d\Gamma^{(L)}_-=d\Gamma\,(1-P_L)/2$, this gives
\begin{equation}
R_{\mu e}(q^2)=\frac{2\,(1-m_\mu^2/q^2)^2}{1-P_L^\mu(q^2)} .
\label{eq:rmuepl}
\end{equation}
This is the same ratio as in Eq.~(\ref{eq:rmuediff}), now written in terms of the muon polarization. In the SM, $R_{\mu e}(q^2)$ therefore carries the same information as $P_L^\mu(q^2)$, but it requires only the rates of the two modes and no analysis of the $\mu$ spin. The $q^2$-integrated form of this relation, $R(1-P_L)=2R_-$, was given for $\bar B\to D\tau\bar\nu$ in Ref.~\cite{Tanaka:2010se}. In particular, at the zero crossing of $P_L^\mu$ the ratio is twice its non-flip part, $R_{\mu e}(q_0^2)=2(1-m_\mu^2/q_0^2)^2=0.35$. For a $q^2$ bin, the same argument gives $\langle P_L^\mu\rangle=1-2\int(1-m_\mu^2/q^2)^2\,d\Gamma_e\big/\!\int d\Gamma_\mu$, with both integrals over the bin.

\begin{figure}[t]
\centering
\includegraphics[width=\textwidth]{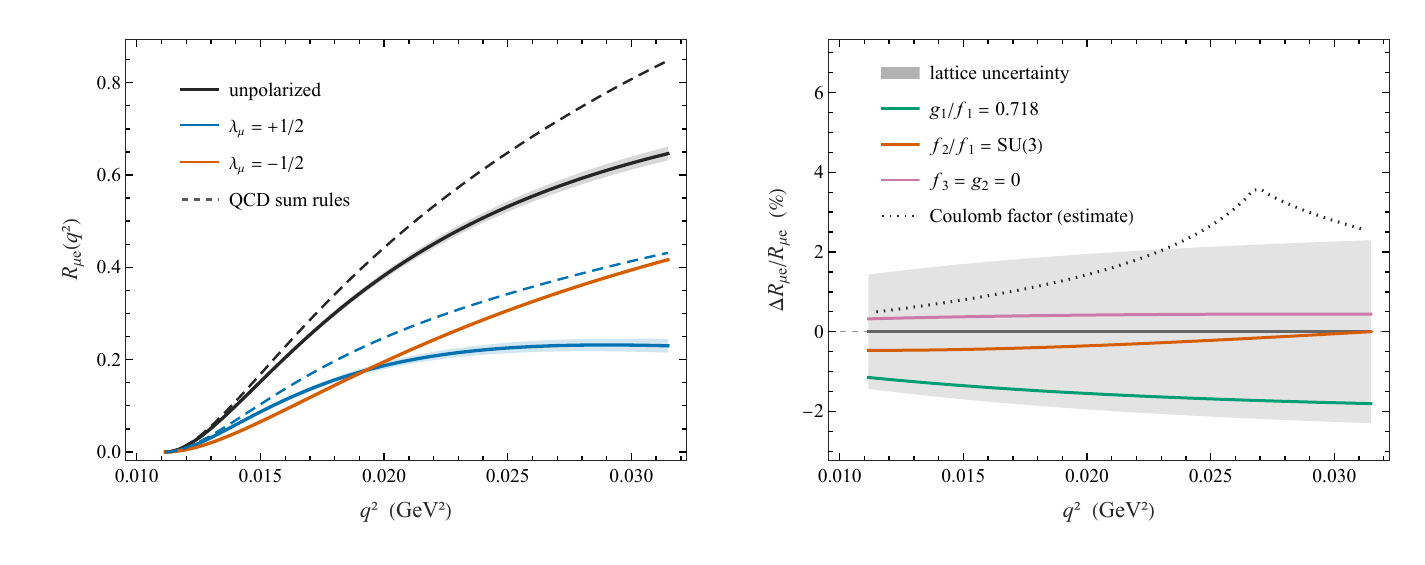}
\caption{Muon-to-electron ratio $R_{\mu e}(q^2)$ with its helicity components (left) and its relative change for different form-factor inputs (right).}
\label{fig:rmue}
\end{figure}

In the left plot of Fig.~\ref{fig:rmue}, we present $R_{\mu e}(q^2)$ and its two helicity components. The $\lambda_\mu=-1/2$ component is given by kinematics alone, Eq.~(\ref{eq:nonflip}). All the dependence on the form factors resides in the $\lambda_\mu=+1/2$ component, i.e., in the ratio of the time-like to the space-like helicity rates, $H_S/(H_U+H_L)$. The two components are equal at $q_0^2$. In Table~\ref{tab:rmuebins}, we give $R_{\mu e}$ in the $q^2$ bins of Table~\ref{tab:bins}, with both modes integrated over the same bin. The last row is the full ratio of Eq.~(\ref{eq:rmue}), with the electron mode integrated from $m_e^2$.

The differential ratio is more sensitive to the form factors than the integrated one, as can be seen in the right plot of Fig.~\ref{fig:rmue}, which shows the relative change of $R_{\mu e}(q^2)$ with respect to the lattice central value when one input is changed and the others are kept at their lattice values. Setting $g_1/f_1$ to the experimental value lowers $R_{\mu e}(q^2)$ by $1.1\%$ at threshold and by $1.8\%$ at $q^2_{\max}$, but leaves the integrated ratio practically unchanged (Sec.~\ref{sec:ffsens}). Up to next-to-leading order in $\delta$, both total rates are proportional to $f_1^2+3g_1^2$~\cite{Chang:2014iba,Kadeer:2005aq}, which is why the integrated ratio does not depend on the form factors~\cite{Chang:2014iba}. The $q^2$ distribution, on the other hand, reflects the lepton-antineutrino angular correlation, which depends on $g_1/f_1$. By Eq.~(\ref{eq:rmuepl}), this is the sensitivity of $P_L^\mu$ to the axial coupling. Replacing the weak magnetism by its SU(3) value or setting $f_3=g_2=0$ changes $R_{\mu e}(q^2)$ by less than $0.5\%$. For the same reason, the lattice uncertainty of $R_{\mu e}(q^2)$, $1.4\%$ at threshold and $2.3\%$ at $q^2_{\max}$, is larger than that of the integrated ratio, $1.1\%$. The QCDSR form factors give a ratio that is up to $31\%$ higher.

We note that Eq.~(\ref{eq:rmuepl}) holds for the tree-level rates. The Coulomb factor discussed in Sec.~\ref{sec:results}\,A depends on $q^2$ for the muon. It raises $R_{\mu e}(q^2)$ by $0.5\%$ near threshold and by up to $3.6\%$ at large $q^2$ (dotted line in Fig.~\ref{fig:rmue} and last column of Table~\ref{tab:rmuebins}). The cusp of the dotted line near $q^2=0.027\GeVsq$ marks the point where the proton and the muon can move with equal velocities. Since it does not depend on the lepton spin, it leaves $P_L^\mu$ unchanged and has to be removed before Eq.~(\ref{eq:rmuepl}) is applied to data. At large $q^2$ it exceeds the form-factor sensitivity, so a complete calculation of the long-distance corrections is required before $R_{\mu e}(q^2)$ can be interpreted at the percent level.

\begin{table}[h]
\caption{$R_{\mu e}$ in $q^2$ bins (in GeV$^2$).}
\label{tab:rmuebins}
\begin{ruledtabular}
\begin{tabular}{lcccc}
bin & lattice & QCDSR (polynomial) & QCDSR ($z$ expansion) & Coulomb factor\\ \hline
$[m_\mu^2,\,0.015]$ & $0.0600\pm0.0009$ & $0.0659$ & $0.0648$ & $+0.7\%$\\
$[0.015,\,0.020]$ & $0.2706\pm0.0048$ & $0.3069$ & $0.2997$ & $+1.1\%$\\
$[0.020,\,0.025]$ & $0.4592\pm0.0093$ & $0.5454$ & $0.5261$ & $+2.0\%$\\
$[0.025,\,q^2_{\max}]$ & $0.5832\pm0.0127$ & $0.7330$ & $0.6972$ & $+3.1\%$\\
$[m_\mu^2,\,q^2_{\max}]$ & $0.3170\pm0.0045$ & $0.3750$ & $0.3518$ & $+2.1\%$\\ \hline
$R_{\mu e}$ (all $q^2$) & $0.1665\pm0.0018$ & $0.1974$ & $0.1786$ & $+2.1\%$\\
\end{tabular}
\end{ruledtabular}
\end{table}

\subsection{Electron channel}
\label{sec:electron}

\begin{figure}[t]
\centering
\includegraphics[width=0.48\textwidth]{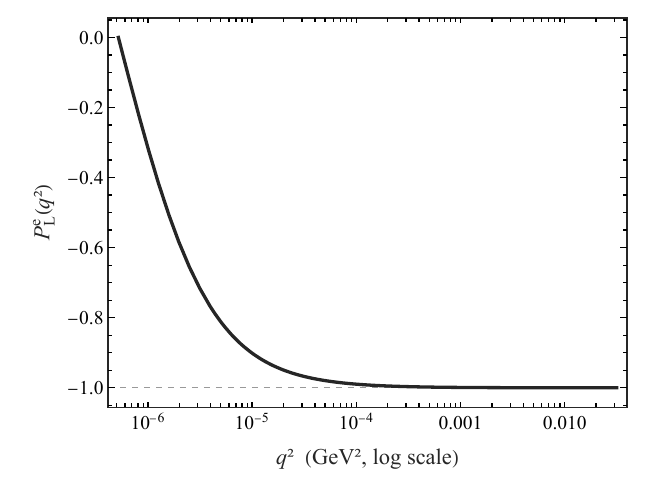}\hfill
\includegraphics[width=0.48\textwidth]{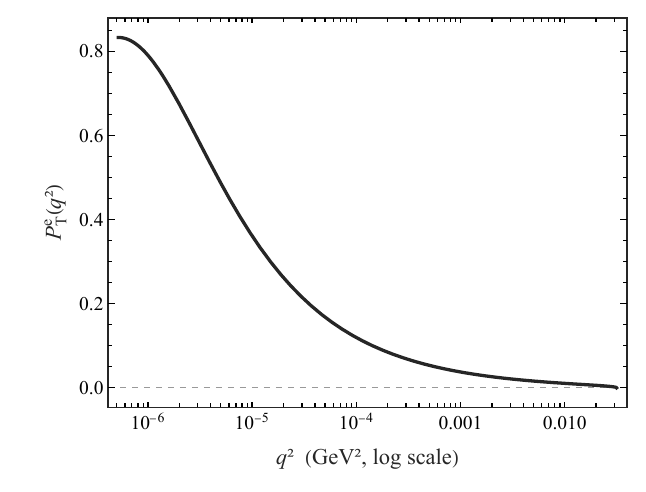}
\caption{Longitudinal (left) and transverse (right) polarization asymmetries of the electron in $\Lpe$.}
\label{fig:electron}
\end{figure}

We now move on to the electron channel, $\Lpe$. The electron is about $200$ times lighter than the muon. The helicity-flip amplitudes, which are suppressed by $m_\ell^2/(2q^2)$, are therefore negligible except very close to the threshold $q^2=m_e^2$, and the electron is produced almost entirely with negative helicity, as expected from the $V-A$ structure of the weak interaction. In Fig.~\ref{fig:electron}, we present the longitudinal (left plot) and transverse (right plot) polarization asymmetries of the electron for the lattice form factors. We use a logarithmic $q^2$ scale, since the polarizations change only at very small $q^2$.

It can be seen from the left plot that $P_L^e$ starts at $+1/3$ at threshold, where the electron is at rest in the dilepton rest frame, and quickly falls to $-1$. It reaches $-0.99$ at $q^2\simeq10^{-4}\GeVsq$, and it equals $-1$ to within $10^{-3}$ for $q^2\gtrsim10^{-3}\GeVsq$. In the same way, $P_T^e$ in the right plot falls from $\simeq0.79$ at threshold to $0.12$ at $q^2=10^{-4}\GeVsq$ and to $0.01$ at $q^2\simeq0.01\GeVsq$, because it is proportional to $m_e/\sqrt{2q^2}$ (Appendix~\ref{app:helicity}). The region $q^2<10^{-4}\GeVsq$, where these mass effects are visible, contains only $0.5\%$ of the rate. The averaged values are therefore close to their massless limits, $\langle P_L^e\rangle=-0.99954$ and $\langle P_T^e\rangle=0.0138\pm0.0003$. The QCDSR form factors give practically the same values, $-0.99954$ and $0.014$. The electron polarizations are thus fixed by the lepton mass and the $V-A$ structure, and they do not test the form factors.

The $A_{\rm FB}$ of the electron, on the other hand, is a sensitive probe of the form factors. For a massless lepton, only the non-flip amplitudes contribute, and the asymmetry is given by the interference of the transverse vector and axial-vector form factors, $\AFB=6\sqrt\lambda\,F_\perp G_\perp/(H_U+H_L)$ (Appendix~\ref{app:helicity}), with $F_\perp=f_1+(1+m_p/m_\Lambda)f_2$ and $G_\perp=g_1-(1-m_p/m_\Lambda)g_2$. We find $\langle\AFB^e\rangle=0.070\pm0.002$ with the lattice form factors, but only $0.006^{+0.004}_{-0.003}$ with the QCDSR form factors, i.e., about twelve times smaller. The reason is the weak magnetism $f_2$. In the lattice calculation, $f_2$ adds to $f_1$ and gives $F_\perp(0)=2.70$. In the QCDSR calculation, $f_2$ has the opposite sign (Table~\ref{tab:couplings}) and almost cancels $f_1$, giving $F_\perp(0)=-0.24$. The axial combination is similar in both cases, $G_\perp(0)=0.83$ and $-0.76$, so the product $F_\perp G_\perp$, and with it the asymmetry, is strongly suppressed. Unlike the muon polarizations, $\langle\AFB^e\rangle$ requires only the measurement of the electron direction and no spin analysis of the lepton. It is therefore a clean experimental probe of the weak magnetism (Sec.~\ref{sec:ffsens}).

\subsection{Observables insensitive and sensitive to the form factors}

We now compare the predictions obtained with the two sets of form factors. The lattice and QCDSR form factors are very different: the QCDSR values of $g_1/f_1$ and $f_2/f_1$ even have the opposite sign to the experimental and lattice values (Table~\ref{tab:couplings}). This comparison therefore shows directly which observables depend on the form factors and which do not. We find that the observables fall into two groups.

The first group consists of observables that are insensitive to the form factors, because they are fixed mainly by the lepton-side helicity structure. For these, the two sets give nearly the same predictions:
\begin{itemize}
\item $\langle\AFB^{+}\rangle_L$: $-0.478$ (lattice) and $-0.477$ (QCDSR). It is fixed by the kinematics of the helicity-flip amplitude, and only the second-class form factors can change it (Sec.~\ref{sec:muonpol});
\item $\langle P_T^\mu\rangle$: $0.53$ (lattice) and $0.57$ (QCDSR), which agree within the QCDSR uncertainty.
\end{itemize}

The second group consists of observables that depend strongly on the form factors. For these, the two predictions differ by much more than the lattice uncertainty:
\begin{itemize}
\item $\langle P_L^\mu\rangle$: $-0.081\pm0.023$ (lattice) and $+0.095$ (QCDSR). With the lattice form factors, $P_L^\mu(q^2)$ changes sign at $q_0^2=0.0193\GeVsq$, whereas with the QCDSR form factors it stays positive. Both the sign of $\langle P_L^\mu\rangle$ and the zero crossing are controlled by the axial coupling $g_1/f_1$;
\item $\langle\AFB^{-}\rangle_L$: $0.096\pm0.003$ and $0.009$, and $\langle\AFB^e\rangle$: $0.070\pm0.002$ and $0.006$. Both are controlled by the weak magnetism $f_2/f_1$, through $F_\perp$ (Sec.~\ref{sec:electron});
\item $R_{\mu e}$: $0.1665\pm0.0018$ and $0.197$. The QCDSR value is larger because its vector form factor $f_1$ rises by $60\%$ between $q^2=0$ and $q^2_{\max}$, compared with $4\%$ for the lattice and with the variation of order $\delta^2$ assumed in Ref.~\cite{Chang:2014iba}.
\end{itemize}

\subsection{Binned predictions}

Experimental results are typically reported after integration over various $q^2$ bins. With this view, we present in Table~\ref{tab:bins} the observables integrated over $q^2$ bins, for comparison with future measurements. For the form-factor-sensitive quantities, the QCDSR central values are given in brackets.

\begin{table}[h]
\caption{Observables integrated over $q^2$ bins (in GeV$^2$).}
\label{tab:bins}
\footnotesize
\begin{ruledtabular}
\begin{tabular}{lcccccc}
\multicolumn{7}{l}{$\Lpmu$}\\
bin & $\Br\times10^5$ & $\langle P_L\rangle$ & $\langle P_T\rangle$ & $\langle\AFB\rangle$ & $\langle\AFB^{+}\rangle_L$ & $\langle\AFB^{-}\rangle_L$\\ \hline
$[m_\mu^2,\,0.015]$ & $0.617\pm0.030$ & $0.182\pm0.013$ [$0.255$] & $0.702\pm0.007$ & $-0.333\pm0.005$ [$-0.397$] & $-0.6276\pm0.0004$ & $0.092\pm0.002$ [$0.008$]\\
$[0.015,\,0.020]$ & $3.18\pm0.17$ & $0.047\pm0.018$ [$0.161$] & $0.656\pm0.010$ & $-0.252\pm0.005$ [$-0.337$] & $-0.5754\pm0.0005$ & $0.104\pm0.003$ [$0.009$]\\
$[0.020,\,0.025]$ & $4.33\pm0.26$ & $-0.090\pm0.023$ [$0.083$] & $0.559\pm0.013$ & $-0.165\pm0.005$ [$-0.266$] & $-0.4891\pm0.0006$ & $0.107\pm0.004$ [$0.010$]\\
$[0.025,\,q^2_{\max}]$ & $4.04\pm0.28$ & $-0.211\pm0.027$ [$0.037$] & $0.369\pm0.012$ & $-0.080\pm0.004$ [$-0.169$] & $-0.3300\pm0.0006$ & $0.082\pm0.004$ [$0.009$]\\ \hline
\multicolumn{7}{l}{$\Lpe$}\\
bin & $\Br\times10^5$ & $\langle P_L\rangle$ & $\langle P_T\rangle$ & $\langle\AFB\rangle$ & & \\ \hline
$[m_e^2,\,0.01]$ & $31.4\pm1.4$ & $-0.99901$ & $0.0234\pm0.0002$ & $0.038\pm0.001$ [$0.003$] & & \\
$[0.01,\,0.02]$ & $25.3\pm1.5$ & $-0.99993$ & $0.0080\pm0.0002$ & $0.093\pm0.002$ [$0.008$] & & \\
$[0.02,\,q^2_{\max}]$ & $16.4\pm1.3$ & $-0.99996$ & $0.0042\pm0.0002$ & $0.097\pm0.004$ [$0.010$] & & \\
\end{tabular}
\end{ruledtabular}
\end{table}

\section{Normal polarization in the Standard Model}
\label{sec:pn}

In this section, we discuss the third component of the lepton polarization vector, $P_N$, which refers to the spin along $\hat e_N=\vecp{\ell}\times\vecp{\Lambda}/|\vecp{\ell}\times\vecp{\Lambda}|$, perpendicular to the decay plane. It measures the triple product $\vec s_\ell\cdot(\vecp{\ell}\times\vecp{\Lambda})$, which is odd under naive time reversal. As noted in Sec.~\ref{sec:theory}, such a term requires a relative phase between amplitudes. At tree level the SM has none, since there is a single amplitude and the form factors are real, so $P_N=0$. A T-odd correlation does not, however, require T violation~\cite{Khriplovich:1967sop}. The two charged particles in the final state, $p$ and $\mu^-$, rescatter electromagnetically, and the absorptive part of this rescattering supplies the phase. In neutral-kaon $K_{\mu3}$ decays, with a charged pion and muon, the resulting muon polarization perpendicular to the decay plane, which is called transverse polarization in Refs.~\cite{Efrosinin:2000yv,Braguta:2002jy} and corresponds to our $P_N$, can be as large as $10^{-3}$~\cite{Okun:1967ww,Efrosinin:2000yv}. In $K^\pm\to\pi^0\mu^\pm\nu$, where the pion is neutral, it is of order $10^{-6}$~\cite{Zhitnitsky:1980he,Efrosinin:2000yv}, and for $B^0\to D^-\mu^+\nu$ it is a few $\times10^{-4}$~\cite{Braguta:2002jy}. The normal polarization of the $\tau$ in $\bar B^0\to D^{(*)}\tau^-\bar\nu_\tau$ has been discussed as a probe of NP in Refs.~\cite{Ivanov:2017mrj,Ajaltouni:2019log}. This effect is the SM baseline of the normal polarization. In this study, we compute it for $\Lpmu$ with the lattice form factors, and we test our method on neutron $\beta$ decay, where the result is known analytically.

\subsection{Method}

At $\mathcal{O}(\alpha)$, only the absorptive part of the one-photon exchange between $p$ and $\mu^-$ contributes to $P_N$. By unitarity~\cite{Cutkosky:1960sp}, it is given exactly by the on-shell rescattering of the final state, as in Ref.~\cite{Braguta:2002jy},
\begin{equation}
\mathcal{M}=\mathcal{M}_W+\frac i2\sum_{\rm spins}\int d\Phi_2(p_1,k_1)\,\mathcal{M}_\gamma(p_1k_1\to p'p_\ell)\,\mathcal{M}_W(\Lambda\to p_1k_1\bar\nu)+\dots,
\label{eq:fsi}
\end{equation}
where $\mathcal{M}_W$ is the tree-level amplitude, Eq.~(\ref{eq:amp}), and $\mathcal{M}_\gamma$ is the one-photon-exchange Born amplitude for $p\,\mu^-\to p\,\mu^-$. For the proton, we use the vertex $\gamma^\mu+i\kappa_p\sigma^{\mu\nu}k_\nu/(2m_p)$ with $\kappa_p=1.793$. The intermediate $p\,\mu^-$ state $(p_1,k_1)$ is integrated over its two-body phase space at fixed total momentum $p'+p_\ell$. The forward region is infrared divergent. There, however, $\mathcal{M}_\gamma\mathcal{M}_W$ tends to a real multiple of $\mathcal{M}_W$ itself, i.e., to a common (Coulomb) phase of the tree amplitude, which cancels in every polarization. We subtract it under the integral, so no photon mass is needed. The dispersive part of the loop is T-even and does not contribute to $P_N$ at this order. $P_N$ then follows from the interference $2\,{\rm Re}[\mathcal{M}_W^*\,(i\,{\rm Abs}\,\mathcal{M})]$ for the two normal spin states.

We evaluate Eq.~(\ref{eq:fsi}) numerically with explicit $4\times4$ Dirac matrices and spinors and the spin projector of Eq.~(\ref{eq:projector}), and sum the intermediate spins with the completeness relations. We have checked that the tree-level $P_N$ vanishes. As a test with a known answer, we applied the same code to neutron $\beta$ decay, $n\to p\,e^-\bar\nu_e$. In the limit of a static (infinitely heavy) proton, the Coulomb attraction between the proton and the electron adds the T-odd term $L\,\vec\xi_e\cdot(\vecp{e}\times\vecp{\nu})/(E_eE_\nu)$ to the decay distribution~\cite{Jackson:1957zz}. Here $\vecp{e}$, $\vecp{\nu}$, $E_e$ and $E_\nu$ are the momenta and energies of the electron and the antineutrino in the neutron rest frame, and $\vec\xi_e$ is the unit vector along the electron spin. This term is the analogue of our $P_N$: it measures the electron spin perpendicular to the plane of $\vecp{e}$ and $\vecp{\nu}$. Its coefficient is $L=\frac{1-g_A^2}{1+3g_A^2}\,\alpha m_e/|\vecp{e}|$, with $g_A$ the axial coupling of the nucleon~\cite{Jackson:1957auh,Ivanov:2018yir}. Our code reproduces $L$ to $0.1\%$, including the sign.

\subsection{Result}

For $\Lpmu$ with the lattice form factors, we find
\begin{equation}
\langle P_N^\mu\rangle_{\rm SM}=-(1.06\pm0.15)\times10^{-3}.
\label{eq:pnsm}
\end{equation}

\begin{figure}[t]
\centering
\includegraphics[width=0.85\textwidth]{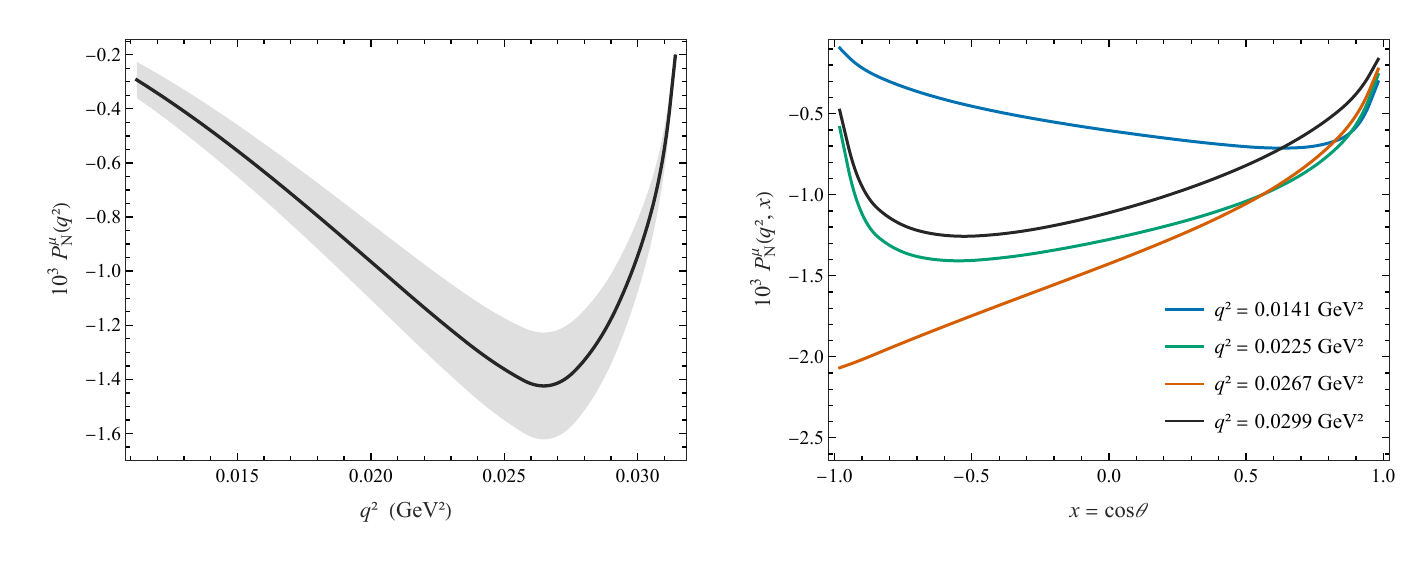}
\caption{Normal polarization of the muon in $\Lpmu$ in the SM as a function of $q^2$ (left) and of $x=\cos\theta$ (right).}
\label{fig:pn}
\end{figure}

In Fig.~\ref{fig:pn}, we show the $q^2$ dependence of this polarization with its uncertainty (left plot) and its dependence on the muon direction, $x=\cos\theta$, for four values of $q^2$ (right plot). It can be seen from the left plot that $|P_N^\mu|$ grows from $3\times10^{-4}$ near threshold to $1.4\times10^{-3}$ at $q^2\simeq0.027\GeVsq$, and vanishes at $q^2_{\max}$, where the decay plane is not defined. The right plot shows that the angular dependence changes with $q^2$: below $q^2\simeq0.02\GeVsq$, where $|P_N^\mu|<10^{-3}$, it is largest for muons emitted forward with respect to the $W$ direction, whereas above it is largest for backward muons and reaches $2\times10^{-3}$ at $q^2\simeq0.027\GeVsq$. These largest values occur where the proton and the muon move slowly relative to each other, so that their electromagnetic attraction acts most strongly. This is also where the Coulomb factor of Sec.~\ref{sec:results}\,A has its cusp (Sec.~\ref{sec:rmueq}).

We estimate the uncertainty from five sources, each propagated to the average over the full phase space. The first three are those of all lattice predictions (Sec.~\ref{sec:lattice}), and the last two concern the photon-proton vertex:
\begin{itemize}
\item setting $g_1/f_1$ to its experimental value changes $\langle P_N^\mu\rangle$ by $-11\%$;
\item the lattice errors of Table~\ref{tab:lattice}, including the $2\%$ systematic uncertainty of the charges, give $8\%$, almost all from $f_1$ and $g_1$; the other four charges and the radii contribute less than $1\%$;
\item scaling all radii by $1.5$ changes it by $+1\%$;
\item including the finite size of the proton through a dipole form factor in the photon-proton vertex changes it by $-3\%$;
\item switching off the anomalous magnetic moment of the proton, $\kappa_p=0$, changes it by $+3\%$.
\end{itemize}
Added in quadrature, these give the $14\%$ uncertainty of Eq.~(\ref{eq:pnsm}). The band in the left plot of Fig.~\ref{fig:pn} shows the same combination at each $q^2$. It decreases from $23\%$ near threshold, where the proton and the muon move fast relative to each other and the structure of the proton matters most, to about $14\%$ at large $q^2$, where the uncertainty of $g_1/f_1$ dominates. In the electron channel, the effect is suppressed by the small electron mass, like the coefficient $L$ in neutron decay, and is of order $10^{-5}$.

With $P_L$, $P_T$ and $P_N$, the full polarization vector of the muon is now predicted. The normal component is two to three orders of magnitude below the in-plane components, as expected for an electromagnetic effect, but it is calculable. It is the SM reference value against which any T-violating contribution to the lepton spin in this decay would have to be compared. Because the FSI conserves CP, the $\ell^+$ in $\bar\Lambda\to\bar p\,\ell^+\nu_\ell$ has the same $P_N$ (Sec.~\ref{sec:theory}), whereas a CP-violating contribution would have opposite signs in the two modes, as in $K^\pm_{\mu3}$ decays~\cite{Efrosinin:2000yv}. Therefore, comparing $\Lambda$ and $\bar\Lambda$ decays separates the two effects, and a difference between the two would reveal CP violation.

\section{Sensitivity to the form factors}
\label{sec:ffsens}

BESIII has measured the axial, weak-magnetism and weak-electricity couplings of $\Lpe$ directly from the decay distribution~\cite{BESIII:2026ehc}. In this section, we therefore identify the hadronic quantity that each observable of Sec.~\ref{sec:results} determines. We vary the five form-factor ratios at $q^2=0$ that are not fixed by the normalization $f_1$: the axial coupling $g_1/f_1$, the weak magnetism $f_2/f_1$, the weak electricity $g_2/f_1$, the induced scalar $f_3/f_1$ and the induced pseudoscalar $g_3/f_1$. We keep $f_1$ and the radii at their lattice values and recompute all observables from our squared amplitudes. The ratios are defined as in Eq.~(\ref{eq:ff}), with $f_2$, $g_2$, $f_3$, $g_3$ normalized by $m_\Lambda$. This is also the convention of Refs.~\cite{Cabibbo:2003cu,BESIII:2026ehc}.

The present knowledge of these ratios is as follows:
\begin{itemize}
\item \emph{Lattice:} $g_1/f_1=0.690$, $f_2/f_1=0.693(17)$, $g_2/f_1=-0.069(16)$, $f_3/f_1=0.052(11)$, $g_3/f_1=-4.34(10)$~\cite{Bacchio:2025auj}.
\item \emph{BESIII:} $g_1/f_1=0.729\pm0.048$ and $f_2/f_1=0.89\pm0.38$ with $g_2=0$, and $g_2/f_1=-0.19\pm0.67$ when all three ratios are fitted~\cite{BESIII:2026ehc}.
\item \emph{SU(3) limit:} conservation of the vector current together with SU(3) symmetry predicts $f_2/f_1=(m_\Lambda/m_p)\,\mu_p/2=1.066$, with $\mu_p$ the anomalous magnetic moment of the proton, and $f_3=0$; in the absence of second-class currents, also $g_2=0$~\cite{Cabibbo:2003cu}.
\end{itemize}
The lattice therefore predicts a sizeable SU(3) breaking of the weak magnetism, which has not yet been tested experimentally.

\begin{table}[h]
\caption{Derivatives of the $q^2$-averaged observables and of $R_{\mu e}$ with respect to the form-factor ratios at $q^2=0$.}
\label{tab:ffsens}
\begin{ruledtabular}
\begin{tabular}{lrrrrr}
Observable & $g_1/f_1$ & $f_2/f_1$ & $g_2/f_1$ & $f_3/f_1$ & $g_3/f_1$\\ \hline
$\langle P_L^\mu\rangle$ & $-0.686$ & $-0.006$ & $0.128$ & $0.070$ & $-0.001$\\
$q_0^2$ (GeV$^2$) & $-0.0190$ & $-0.0003$ & $0.0035$ & $0.0020$ & $-0.00005$\\
$\langle P_T^\mu\rangle$ & $-0.362$ & $0.044$ & $0.087$ & $0.001$ & $-0.002$\\
$\langle\AFB^\mu\rangle$ & $0.150$ & $0.043$ & $-0.041$ & $-0.005$ & $0.001$\\
$\langle\AFB^{+}\rangle_L$ & $-0.003$ & $0.000$ & $-0.022$ & $0.033$ & $0.001$\\
$\langle\AFB^{-}\rangle_L$ & $-0.085$ & $0.077$ & $0.011$ & $0$ & $0$\\
$\langle\AFB^e\rangle$ & $-0.018$ & $0.056$ & $-0.003$ & $0$ & $0$\\
$R_{\mu e}$ & $-0.0021$ & $-0.0009$ & $-0.0080$ & $0.0109$ & $-0.0002$\\
\end{tabular}
\end{ruledtabular}
\end{table}

\begin{figure}[t]
\centering
\includegraphics[width=\textwidth]{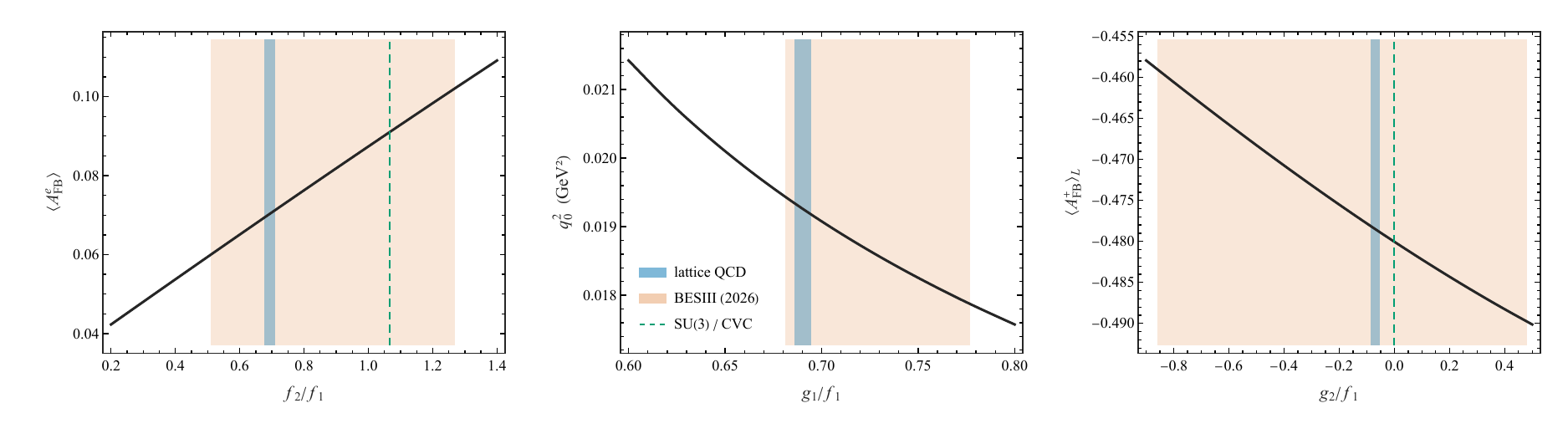}
\caption{Dependence of $\langle\AFB^e\rangle$ on $f_2/f_1$ (left), of $q_0^2$ on $g_1/f_1$ (middle) and of $\langle\AFB^{+}\rangle_L$ on $g_2/f_1$ (right).}
\label{fig:ffsens}
\end{figure}

Table~\ref{tab:ffsens} gives the derivatives of the observables with respect to each ratio, evaluated at the lattice values with the other ratios fixed. Here $q_0^2$ is the value of $q^2$ at which $P_L^\mu(q^2)$ changes sign (Sec.~\ref{sec:muonpol}). In Fig.~\ref{fig:ffsens}, we show three observables as functions of the form factor that each mainly measures, for the three ratios measured by BESIII, with the other ratios fixed at their lattice values. The solid curves are our predictions, the bands show the lattice and BESIII values listed above, and the dashed lines the SU(3) limit. Where a curve crosses a band or a line, it gives the value of the observable predicted for that input. Read in the opposite direction, each curve turns a measurement of the observable into a value of the ratio, and the steeper the curve, the more sensitive the observable is to that ratio. From Table~\ref{tab:ffsens} and Fig.~\ref{fig:ffsens}, one can see four clear patterns.

\emph{(i) The longitudinal polarization measures the axial coupling.} $\langle P_L^\mu\rangle$ and its zero crossing depend essentially only on $g_1/f_1$: $q_0^2$ moves by $-0.019\GeVsq$ per unit of $g_1/f_1$. This is why the QCDSR form factors, with $g_1/f_1<0$, give $P_L^\mu>0$ without a zero crossing.

\emph{(ii) The non-flip asymmetries measure the weak magnetism.} For $m_\ell\to0$, Appendix~\ref{app:helicity} gives $\AFB\propto F_\perp G_\perp/(H_U+H_L)$ with $F_\perp=f_1+(1+m_p/m_\Lambda)f_2=f_1(1+1.84\,f_2/f_1)$. A change of $g_1/f_1$ enters both the numerator and the denominator and largely cancels, whereas $f_2/f_1$ enters through $F_\perp$, so $\langle\AFB^e\rangle$ and $\langle\AFB^{-}\rangle_L$ respond mainly to the weak magnetism (Table~\ref{tab:ffsens}). Changing $f_2/f_1$ from the lattice value to the SU(3) value increases $\langle\AFB^e\rangle$ from $0.070$ to $0.091$, i.e., by $30\%$.

\emph{(iii) The helicity-flip asymmetry measures the second-class form factors.} $\langle\AFB^{+}\rangle_L$ is insensitive to $g_1/f_1$ and $f_2/f_1$. This is why it is the least sensitive to the form factors of all the predictions of Sec.~\ref{sec:results}. Its deviation from the kinematic value is driven only by the second-class form factors, $-0.022\,g_2/f_1+0.033\,f_3/f_1$. It is therefore a clean, if experimentally demanding, probe of these form factors: the present BESIII range of $g_2/f_1$ corresponds to $\pm0.015$.

\emph{(iv) The muon-to-electron ratio is nearly free of the leading form factors.} In $R_{\mu e}$, the dependence on $g_1/f_1$ and $f_2/f_1$ largely cancels between numerator and denominator. Per unit change of the ratios, $R_{\mu e}$ changes by $-1.3\%$ and $-0.5\%$, respectively. Replacing the lattice weak magnetism by its SU(3) value changes $R_{\mu e}$ by only $0.2\%$, compared with $30\%$ for $\langle\AFB^e\rangle$. Even when $g_1/f_1$ and $f_2/f_1$ are both replaced by the QCDSR values, which have the opposite sign, $R_{\mu e}$ changes by only $0.5\%$. Among the ratios at $q^2=0$, the residual dependence comes from the second-class form factors, whose lattice uncertainties change $R_{\mu e}$ by only about $0.1\%$. As far as the hadronic input is concerned, $R_{\mu e}$ is therefore a clean SM prediction.

Of these observables, the electron asymmetry is the easiest to measure now, because it needs only the direction of the electron and no measurement of its spin. Since $\langle\AFB^e\rangle$ changes by $0.056$ per unit of $f_2/f_1$ (Table~\ref{tab:ffsens}), a measurement of it directly determines the weak magnetism. For an asymmetry $A$ measured with $N$ events, the statistical uncertainty is $\sigma_A\simeq\sqrt{(1-A^2)/N}$. With the $1791$ signal and $207$ background events selected by BESIII for the form-factor fit~\cite{BESIII:2026ehc}, we estimate $\sigma(\AFB^e)\simeq0.025$, including the dilution by the background. At the STCF, $3.4\times10^{12}$ $J/\psi$ per year and $104$ reconstructed decays per $10^9$ $J/\psi$~\cite{Zhou:2025skh} give about $3.5\times10^5$ decays per year and hence, neglecting backgrounds, $\sigma(\AFB^e)\simeq1.7\times10^{-3}$. This would determine $f_2/f_1$ to $\pm0.03$ and separate the lattice and SU(3) values of the weak magnetism at the level of $12\sigma$, or $10\sigma$ when the theoretical uncertainty of the other form factors is included.

\section{conclusions}
\label{sec:concl}

In this work, we have studied the polarization of the charged lepton in the semileptonic hyperon decay $\Lp$ ($\ell=e,\mu$), which is induced by the quark-level transition $s\to u$. This decay channel provides a framework for testing the SM and the SU(3) breaking in the baryonic form factors. In particular, we investigated the complete polarization vector of the charged lepton within the SM. Using the lattice-QCD form factors of Ref.~\cite{Bacchio:2025auj}, including their statistical and systematic uncertainties, we calculated the $q^2$-dependent observables, which include the differential branching ratio, the longitudinal, transverse and normal polarization asymmetries, and the forward-backward asymmetry of each lepton spin state. For comparison, we also used the QCD sum-rule form factors of Ref.~\cite{Ahmadi:2025oal}.

Our analysis shows that, because the phase space of $\Lpmu$ is narrow ($q^2_{\max}\simeq2.8\,m_\mu^2$), the two muon helicities contribute comparably. As a result, the longitudinal polarization changes sign at $q_0^2=0.0193\pm0.0006\GeVsq$ and averages to $\langle P_L^\mu\rangle=-0.081\pm0.023$, while the transverse polarization is large and precisely predicted, $\langle P_T^\mu\rangle=0.529\pm0.012$. We also computed the T-odd normal polarization, generated by the electromagnetic FSI of the proton and the muon, from the absorptive part of one-photon exchange. We obtained $\langle P_N^\mu\rangle=-(1.06\pm0.15)\times10^{-3}$. This result complements our predictions for $P_L^\mu$ and $P_T^\mu$ and provides the reference value for future tests of T violation with the lepton spin.

Furthermore, we have shown that the observables separate according to the hadronic quantity they measure. $\langle P_L^\mu\rangle$ and its zero crossing measure $g_1/f_1$. The electron asymmetry, $\langle\AFB^e\rangle=0.070\pm0.002$, and the asymmetry of negative-helicity muons, $\langle\AFB^{-}\rangle_L=0.096\pm0.003$, measure the weak magnetism $f_2/f_1$, given the precisely known $g_1/f_1$, whereas the asymmetry of positive-helicity muons, $\langle\AFB^{+}\rangle_L=-0.4785\pm0.0008$, is fixed by the lepton-side kinematics up to a dependence on the second-class form factors $g_2$ and $f_3$ only. At the STCF, a measurement of $\langle\AFB^e\rangle$ would distinguish the lattice and SU(3) values of the weak magnetism at the level of $10\sigma$. The QCD sum-rule form factors of Ref.~\cite{Ahmadi:2025oal}, whose $g_1/f_1$ and $f_2/f_1$ disagree in sign with experiment and lattice QCD, give very different values for the form-factor-sensitive observables ($\langle P_L^\mu\rangle>0$, no zero crossing, $\langle\AFB^{-}\rangle_L\simeq0.01$, $\langle\AFB^e\rangle\simeq0.006$), while they agree on $\langle P_T^\mu\rangle$ and, through a cancellation between their second-class and other form factors (Sec.~\ref{sec:muonpol}), on $\langle\AFB^{+}\rangle_L$.

In addition, we computed the LFU ratio and obtained $R_{\mu e}=0.1665\pm0.0018$, which is nearly independent of $g_1/f_1$ and $f_2/f_1$; its residual hadronic dependence comes mainly from the $q^2$ dependence of the form factors. The Coulomb attraction of the slow proton and muon alone raises $R_{\mu e}$ by about $2\%$, and the emission of real photons partly cancels this increase, so we expect a net correction of order $1\%$ to $2\%$. We have also shown that, in the SM, the $q^2$ dependence of $R_{\mu e}$ is fixed by $P_L^\mu(q^2)$, so that $P_L^\mu$, including its sign change, can be measured from the rates of the $e$ and $\mu$ modes, once these rates are corrected for the electromagnetic effects mentioned above.

Overall, our results demonstrate that the lepton polarization observables complement the branching ratios and $R_{\mu e}$ as tests of the $\Lambda\to p$ form factors and of the SM description of the $s\to u$ transition. The normal polarization is also sensitive to CP-violating scalar, pseudoscalar and tensor interactions. Lattice form factors for the scalar and tensor interactions have recently become available~\cite{Alexandrou:2026noh}. We will discuss these NP effects in a separate work.

\begin{acknowledgments}
Q.~M.~U.~Salam is very thankful to the National Center for Physics (NCP), Islamabad, where he is a hosted researcher, for its hospitality during his stay while this manuscript was written.
\end{acknowledgments}

\appendix
\section{Helicity-amplitude expressions}
\label{app:helicity}

In the helicity formalism~\cite{Korner:1989qb,Korner:1994nh,Gutsche:2014zna,Kadeer:2005aq}, the unpolarized rate reads
\begin{equation}
\frac{d\Gamma}{dq^2}=\frac{G_F^2S_{\rm EW}|V_{us}|^2}{(2\pi)^3}\,\frac{\lambda^{1/2}(q^2-m_\ell^2)^2}{48\,m_\Lambda^3\,q^2}
\Big[(H_U+H_L)(1+\delta_\ell)+3\delta_\ell H_S\Big],\qquad \delta_\ell\equiv\frac{m_\ell^2}{2q^2},
\label{eq:helrate}
\end{equation}
with $H_U=|H_{\frac12,1}|^2+|H_{-\frac12,-1}|^2$, $H_L=|H_{\frac12,0}|^2+|H_{-\frac12,0}|^2$ and $H_S=|H_{\frac12,t}|^2+|H_{-\frac12,t}|^2$, where $H_{\lambda_p,\lambda_W}=H^V_{\lambda_p,\lambda_W}-H^A_{\lambda_p,\lambda_W}$. In terms of the form factors $F_i$, $G_i$ of Eq.~(\ref{eq:basis}), the helicity amplitudes read~\cite{Ahmadi:2025oal}, with $\sigma=(m_\Lambda^2+m_p^2-q^2)/(2m_\Lambda m_p)$,
\begin{align}
H^V_{\frac12,0}&=\sqrt{\frac{2m_\Lambda m_p(\sigma-1)}{q^2}}\Big[(m_\Lambda+m_p)F_1+m_p(\sigma+1)F_2+m_\Lambda(\sigma+1)F_3\Big],\nonumber\\
H^A_{\frac12,0}&=\sqrt{\frac{2m_\Lambda m_p(\sigma+1)}{q^2}}\Big[(m_\Lambda-m_p)G_1-m_p(\sigma-1)G_2-m_\Lambda(\sigma-1)G_3\Big],\nonumber\\
H^V_{\frac12,1}&=-2\sqrt{m_\Lambda m_p(\sigma-1)}\,F_1,\qquad H^A_{\frac12,1}=-2\sqrt{m_\Lambda m_p(\sigma+1)}\,G_1,\nonumber\\
H^V_{\frac12,t}&=\sqrt{\frac{2m_\Lambda m_p(\sigma+1)}{q^2}}\Big[(m_\Lambda-m_p)F_1+(m_\Lambda-m_p\sigma)F_2+(m_\Lambda\sigma-m_p)F_3\Big],\nonumber\\
H^A_{\frac12,t}&=\sqrt{\frac{2m_\Lambda m_p(\sigma-1)}{q^2}}\Big[(m_\Lambda+m_p)G_1-(m_\Lambda-m_p\sigma)G_2-(m_\Lambda\sigma-m_p)G_3\Big],
\label{eq:helamp}
\end{align}
and $H^V_{-\lambda_p,-\lambda_W}=H^V_{\lambda_p,\lambda_W}$, $H^A_{-\lambda_p,-\lambda_W}=-H^A_{\lambda_p,\lambda_W}$. Besides $H_U$, $H_L$ and $H_S$, we need the parity-odd and the scalar-longitudinal combinations
\begin{equation}
H_P=|H_{\frac12,1}|^2-|H_{-\frac12,-1}|^2=-8\sqrt{\lambda}\,F_\perp G_\perp,\qquad
H_{SL}=H_{\frac12,0}H_{\frac12,t}+H_{-\frac12,0}H_{-\frac12,t},
\label{eq:HP}
\end{equation}
where $F_\perp\equiv F_1=f_1+(1+r)f_2$ and $G_\perp\equiv G_1=g_1-(1-r)g_2$ are the transverse helicity form factors~\cite{Bacchio:2025auj}, and $H_{\rm tot}=(H_U+H_L)(1+\delta_\ell)+3\delta_\ell H_S$.

With $x=\cos\theta$ as defined in Sec.~\ref{sec:theory}, the distributions for the two lepton helicities (longitudinal spin projection $k=-1$: non-flip; $k=+1$: helicity flip) are
\begin{align}
\frac{d^2\Gamma^{(L)}_{-}}{dq^2dx}&=\frac{3}{2}\frac{d\Gamma/dq^2}{H_{\rm tot}}\Big[\frac{1+x^2}{4}H_U+\frac{1-x^2}{2}H_L-\frac{x}{2}H_P\Big],\nonumber\\
\frac{d^2\Gamma^{(L)}_{+}}{dq^2dx}&=\frac{3}{2}\frac{d\Gamma/dq^2}{H_{\rm tot}}\,\delta_\ell\Big[\frac{1-x^2}{2}H_U+x^2H_L+H_S-2x\,H_{SL}\Big],
\end{align}
and for spin projection $k=\pm1$ along $\hat e_T$
\begin{equation}
\frac{d^2\Gamma^{(T)}_{\pm}}{dq^2dx}=\frac12\sum_{k=\pm}\frac{d^2\Gamma^{(L)}_{k}}{dq^2dx}\pm\frac{3}{2}\frac{d\Gamma/dq^2}{H_{\rm tot}}\sqrt{\delta_\ell}\,\frac{\sqrt{1-x^2}}{\sqrt2}\Big[x\Big(\frac{H_U}{2}-H_L\Big)-\frac{H_P}{2}+H_{SL}\Big].
\end{equation}
Integrating over $x$ gives
\begin{align}
P_L(q^2)&=\frac{\delta_\ell\,(H_U+H_L+3H_S)-(H_U+H_L)}{H_{\rm tot}},\qquad
P_T(q^2)=-\frac{3\pi\sqrt{\delta_\ell}}{4\sqrt2}\,\frac{H_P-2H_{SL}}{H_{\rm tot}},\nonumber\\
\AFB(q^2)&=-\frac34\,\frac{H_P+4\delta_\ell H_{SL}}{H_{\rm tot}},\qquad
\AFB^{(L),-}(q^2)=-\frac34\,\frac{H_P}{H_U+H_L},\qquad
\AFB^{(L),+}(q^2)=-\frac{3H_{SL}}{H_U+H_L+3H_S},\nonumber\\
\AFB^{(T),\pm}(q^2)&=\frac{-\frac14 H_P-\delta_\ell H_{SL}\pm\frac{\sqrt2}{3}\sqrt{\delta_\ell}\,\big(\frac{H_U}{2}-H_L\big)}{\frac13 H_{\rm tot}\mp\frac{\pi}{2\sqrt2}\sqrt{\delta_\ell}\,\big(\frac{H_P}{2}-H_{SL}\big)}.
\label{eq:closed}
\end{align}
The signs refer to our choices of $\theta$ and $\hat e_T$, and of the phases in Eq.~(\ref{eq:helamp}). These expressions agree exactly with our full squared amplitudes, for both leptons, at all $q^2$ and for several sets of form factors.

\section{QCD sum-rule parametrizations}
\label{app:qcdsr}

The QCDSR form factors $F_i$ and $G_i$ of Ref.~\cite{Ahmadi:2025oal}, related to ours by Eq.~(\ref{eq:basis}), are given with two parametrizations of their $q^2$ dependence. Denoting any of them by $\mathcal{F}$, the polynomial fit is
\begin{equation}
\mathcal{F}(q^2)=\frac{\mathcal{F}(0)}{1-a_1x+a_2x^2+a_3x^3+a_4x^4},\qquad x=\frac{q^2}{m_\Lambda^2},
\label{eq:qcdsrpoly}
\end{equation}
and the $z$ expansion is
\begin{equation}
\mathcal{F}(q^2)=\frac{\mathcal{F}(0)+A_1\big[z(q^2)-z(0)\big]}{1-q^2/m^2},\qquad z(q^2)=\frac{\sqrt{t_+-q^2}-\sqrt{t_+-t_0}}{\sqrt{t_+-q^2}+\sqrt{t_+-t_0}},
\label{eq:qcdsrz}
\end{equation}
with $t_\pm=(m_\Lambda\pm m_p)^2$ and $t_0=t_+\big(1-\sqrt{1-t_-/t_+}\big)$. The pole mass is $m=0.892$~GeV ($K^*$) for $F_i$ and $m=1.27$~GeV ($K_1$) for $G_i$~\cite{Ahmadi:2025oal}. The values of $\mathcal{F}(0)$ are given in Sec.~\ref{sec:qcdsr}, and the coefficients $a_k$ and $A_1$ are listed in Tables~II and III of Ref.~\cite{Ahmadi:2025oal}.

\end{document}